\documentclass[final]{llncs}
\usepackage[utf8]{inputenc}  
\usepackage[colorlinks=false,breaklinks=true]{hyperref}
\usepackage{eepic}
\usepackage{etex}
\usepackage{pictex}
\usepackage{fixme}
\usepackage{color}
\hypersetup{
    colorlinks,%
    citecolor=black,%
    filecolor=black,%
    linkcolor=black,%
    urlcolor=black
}

\usepackage{graphicx,subfigure}
\usepackage{amssymb}

\def\fullversion{fullversion}

\newcommand{\rmax}{\mathrm{RMAX}}
\newcommand{\rmin}{\mathrm{RMIN}}
\newcommand{\thresh}{{T}}
\newcommand{\ngb}{N} 

\newcommand{\ngbl}{\ngb^{<}}
\newcommand{\ngbg}{\ngb^{>}}

\newcommand{\transiop}{\mathrm{TP}}
\newcommand{\separmin}{\Delta^{\wedge}}
\newcommand{\ewg}[2][]{G_{#1}[{#2}]}

\newcommand{\origin}{\mathbf{o}}
\newcommand{\hmtop}{\mathrm{HMT}}
 
\newcommand{\apixp}{{p}}  
\newcommand{\apixq}{{q}}  

\newcommand{\apath}{{\mathcal P}}

\newcommand{\range}{{\mathsf{R}}} 

\newcommand{\arangeg}{\omega} 
\newcommand{\arangel}{\alpha} 

\newcommand{\locglops}{(\arangepsl,\arangepsg)}
\newcommand{\arangepsg}{\arangeg} 
\newcommand{\arangepsl}{\arangel} 

\newcommand{\RCC}{{\mathrm{\mbox{-}CC}}}

\newcommand{\diss}{d^{\star}}
\newcommand{\dissloc}{\diss_{A}}
\newcommand{\dissglo}{\diss_{\Omega}}

\newcommand{\umetric}{d}
\newcommand{\acluster}{C}  

\newcommand{\dissalpha}{\alpha^\star}

\newcommand{\alphadegree}{\dissalpha\mathrm{\mbox{-}deg}}
\newcommand{\alphadegreeval}[1]{#1\mathrm{\mbox{-}deg}}
\newcommand{\alphaparam}{n}

\newcommand{\st}{\; | \;}
\newcommand{\elimine}[1]{}
\newcommand{\Bar}[1]{\overline{#1}}

\renewcommand{\refname}{References}

\newcommand\citep[1]{\cite{#1}}
\newcommand\citet[1]{\cite{#1}}
\begin{document}

\frontmatter

\addtocmark{On morphological hierarchical representations}

\mainmatter

\title{On morphological hierarchical representations\\ for image processing and spatial data clustering\thanks{A preliminary version of this paper was presented at the workshop WADGMM~2010~\citep{koethe-montanvert-soille2010} held in conjunction with ICPR~2010, Istanbul, August 2010.}}

\titlerunning{On morphological hierarchical representation}

\author{Pierre Soille$^{1}$ and Laurent Najman$^{2}$}
\authorrunning{Laurent Najman and Pierre Soille}
\tocauthor{Laurent Najman, Pierre Soille}

\institute{
$^{1}$Institute for the Protection and Security of the Citizen,\\
Joint Research Centre, European Commission,\\
Via E. Fermi 2749, I-21027 Ispra (VA), Italy\\
$^{2}$Universit\'e Paris-Est, Laboratoire d'Informatique Gaspard-Monge\\
 Equipe A3SI, ESIIE, France\\}

\maketitle


\begin{abstract}
Hierarchical data representations in the context of classification and data clustering were put forward during the fifties. Recently, hierarchical image representations have gained renewed interest for segmentation purposes. In this paper, we briefly survey fundamental results on hierarchical clustering and then detail recent paradigms developed for the hierarchical representation of images in the framework of mathematical morphology: constrained connectivity and ultrametric watersheds. Constrained connectivity can be viewed as a way to constrain an initial hierarchy in such a way that a set of desired constraints are satisfied. The framework of ultrametric watersheds provides a generic scheme for computing any hierarchical connected clustering, in particular when such a hierarchy is constrained.
The suitability of this framework for solving practical problems is illustrated with applications in remote sensing.\\


\end{abstract}

\ifdefined\fullversion
\fi

\section{Introduction}

Most image processing applications require the selection of an image representation suitable for further analysis.  The suitability of a given representation can be evaluated by confronting its properties with those required by the application at hand.  In practice, images are often represented by decomposing them into primitive or fundamental elements that can be more easily interpreted. Examples of decomposition (or simply representation) schemes are given hereafter:
\begin{itemize}
\item A functional decomposition decomposes the image into a sum of elementary functions.  The most famous functional decomposition is the Fourier transform which decomposes the image into a sum of cosine functions with a given frequency, phase, and amplitude.  This proves to be a very effective  representation for applications requiring to target structures corresponding to well-defined frequencies;
\item A pyramid decomposition relies on a shrinking operation which applies a low-pass filter to the image and downsamples it by a factor of two and an expand operation which upsamples the image by a factor of two using a predefined interpolation method.  Such a scheme is extremely efficient in situations where the analysis can be initiated at a coarse resolution and refined by going through levels of increasing resolution;
\item A multi-scale representation consists of a one-parameter family of filtered images, the parameter indicating the degree (scale) of filtering.  This scheme is appropriate for the analysis of complex images containing structures at various scales;
\item A skeleton representation consists in representing the image by a thinned version.  It is useful for applications where the geometric and topological properties of the image structures need to be measured;
\item The threshold decomposition decomposes a grey tone image into a stack of binary images corresponding to its successive threshold levels.  This decomposition is useful as a basis for some hierarchical representations (see below) and from a theoretical point of view for generalising operations on binary images to grey tone images;
\item A hierarchical representation of an image can be viewed as an ordered set or tree (acyclic graph) with some elementary components defining its leaves and the full image domain defining its root.  Examples of elementary components are the regional minima/maxima/extrema, or the flat zones of the input image.  This approach is interesting in all applications where the tree encoding the hierarchy offers a suitable basis for revealing structural information for filtering or segmentation purposes.
\end{itemize}
Note that these schemes are not mutually exclusive.  A case in point is the skeleton representation defined in terms of maximal inscribed disks since it fits the multi-scale representation (with morphological openings with disks of increasing size as structuring elements) as well as the functional decomposition (with spatially localised disks as elementary functions that are unioned to reconstruct the original pattern).

A given representation scheme can be further characterised by considering the properties of the operations it relies on.  For example, a representation is linear if it is based on operations invariant to linear transformations of the input image.  The multi-scale representation with Gaussian filters of increasing size fulfils this property.  Morphological representations are non-linear representations relying on morphological operations.  For example, a granulometry is a morphological multi-scale representation originally proposed by Matheron in his seminal study on the analysis of porous media~\citep{matheron67}.  
The representation does not need to rely exclusively on morphological operations to be considered as morphological.  For example, the non-linear scale-space representation with levellings~\citep{meyer-maragos2000} is based on self-dual geodesic reconstruction using Gaussian filters of increasing size as geodesic mask.

This paper deliberately focuses on hierarchical image representations for image segmentation with emphasis on morphological methods.  Note that the development of hierarchical representations appeared first in taxonomy in the form of hierarchical clustering methods (see for example \citep{cormack71} for an old but excellent review on classification including a discussion on hierarchical clustering).  In fact, hierarchical image segmentation can be seen as a hierarchical clustering of spatial data.
Graph theory is the correct setting for formalising clustering concepts as already recognised in \citep{estabrook66} and \citep{matula70}, see also the enlightening paper~\citep{zahn71} as well as the detailed survey and connections between graph theory and clustering in \citep{hubert74} (and \citep{hubert73} for clustering on directed graphs).  For this reason, Sec.~\ref{s.graph} presents briefly background notions and notations of graph theory used throughout this paper.  Then, fundamental concepts of hierarchical clustering methods where the spatial location of the data points is usually not taken into account are reviewed in  Sec.~\ref{s.hclustering}.  Hierarchical image segmentation methods where the spatial location of the observations (i.e., the pixels) plays a central role are presented in a nutshell in Sec.~\ref{s.his}.  Recent recent paradigms developed for the hierarchical representation of images in the framework of mathematical morphology known as constrained connectivity and ultrametric watersheds are then developed in Sec.~\ref{s.ccuw} while highlighting their links with hierarchical clustering methods.  The framework of ultrametric watersheds provides a generic scheme for computing any hierarchical connected clustering, in particular when such a hierarchy is constrained.  Before concluding, the problem of transition pixels is set forth in Sec.~\ref{s.transp}.

\section{Background definitions and notations on graphs}\label{s.graph}
The objects under study (specimens in biology, galaxies in astronomy, or pixels in image processing) are considered as the nodes of a graph.  An edge is then drawn between all pairs of objects that need to be compared.  The comparison often relies on a dissimilarity measure that assigns a weight to each edge.  Following the notations of \citep{Diestel97}, we summarise hereafter graph definitions required in the context of clustering. 

A {\em graph} is defined as a pair~$X = (V,E)$ where~$V$ is a finite
set and~$E$ is composed of unordered pairs of~$V$, {\em i.e.},~$E$ is
a subset of~$\left\{\{p,q\} \subseteq V \st p\neq q \right\}$. 
Each element of~$V$ is called a {\em vertex or a point (of~$X$)}, and
each element of~$E$ is called an {\em edge (of $X$)}. If~$V \neq
\emptyset$, we say that~$X$ is {\em non-empty}.

As several graphs are considered in this paper, whenever this is
necessary, we denote by $V(X)$ and by $E(X)$ the vertex and edge set
of a graph $X$.

Let~$X$ be a graph. If~$u = \{p,q\}$ is an edge of~$X$, we say
that~$p$ and~$q$ are {\em adjacent (for~$X$)}. Let~$\pi=\langle p_0,
\ldots , p_\ell \rangle$ be an ordered sequence of vertices
of~$X$,~$\pi$ is {\em a path from~$p_0$ to~$p_\ell$ in~$X$ (or
in~$V$)} if for any~$i \in [1,\ell]$,~$p_i$ is adjacent to
$p_{i-1}$. In this case, we say that {\em~$p_0$ and~$p_\ell$ are
linked for~$X$}.
We say that $X$ {\em is connected\/} if any two vertices of~$X$ are
linked for~$X$.

Let~$X$ and~$Y$ be two graphs. If~$V(Y) \subseteq V(X)$ and~$E(Y)
\subseteq E(X)$, we say that {\em~$Y$ is a subgraph of~$X$} and we
write~$Y \subseteq X$. We say that~$Y$ is a {\em
connected component of~$X$}, or simply a {\em component of~$X$},
if~$Y$ is a connected subgraph of~$X$ which is maximal for this
property, {\em i.e.}, for any connected graph~$Z$,~$Y \subseteq Z
\subseteq X$ implies~$Z = Y$.

Clustering methods generally work on a complete graph $(V,V\times
V)$. In this case, the notion of connected component is not an
important one, as any subset is obviously connected. On contrary, this
notion is fundamental for image segmentation.

Let $X$ be a graph, and let~$S \subseteq E(X)$. The {\em graph induced
by $S$} is the graph whose edge set is~$S$ and whose vertex set is
made of all points that belong to an edge in~$S$, {\em i.e.},~$(\{p
\in V(X) \st \exists u \in S, p \in u\}, S)$.

In the sequel of this paper, $X=(V,E)$ denotes a connected graph, and the letter $V$ ({\em resp.} $E$) will always  refer to the vertex set ({\em resp.} the edge set) of $X$. We will also assume that~$E \neq \emptyset$.  Let $S\subset E$. In the following, when no confusion may occur, the graph induced by~$S$ is also denoted by~$S$.  If~$S\subset E$, we denote by {\em~$\Bar{S}$ the complementary set of~$S$ in~$E$}, i.e.,~$\Bar{S} = E \setminus S$.

Typically, in applications to image segmentation,~$V$ is the set of
picture elements (pixels) and~$E$ is any of the usual adjacency
relations, {\em e.g.}, the 4- or 8-adjacency in 2D~\citep{KR89}.
In all examples, 4-adjacency is used.

We consider in this paper weighted graphs, and either the vertices or
the edges of a graph can be weighted. We denote the weight on the
vertives of $V$ by $f$, and the weights on the edges of $E$ by $F$.  For
application to image processing, $f$ is generally some information on
the pixels (e.g., the grey level of the considered pixel), and $F$
represents a dissimilarity (e.g., $F(\{p,q\}) = |f(p)-f(q)|$).

\section{Hierarchical clustering}\label{s.hclustering}

Clustering can be defined as a method for grouping objects into homogeneous groups (called clusters) on the basis of empirical measures of similarity among those objects.  Ideally, the method should generate clusters maximising their internal cohesion and external isolation.  
Analogously to the categorisation of classification methods proposed in~\citep{spaerckjones70}, any clustering methodology can be characterised by three main properties.  The first concerns the relation between object properties and clusters.  It indicates whether the clusters are monothetic or polythetic.  A cluster is monothetic if and only if all its members share the same common property or properties.  The second property regards the relation between objects and clusters.  It indicates whether the clusters are exclusive (i.e., non-overlapping) or overlapping.  Non-overlapping clustering methods can be defined as partitional in the sense that they realise a partition of the input objects (a partition of a set is defined as division of this set in disjoint non-empty subsets such that their union is equal to this set).   Non-partitional clustering allows for overlap between clusters, see~\citep{jardine-sibson68} for an early reference on this topic and~\citep{barthelemy-brucker-osswald2004} for recent developments.  The third property refers to the relation between clusters.  It indicates whether the clustering method is hierarchical (also called ordered) or non-hierarchical (unordered).

Because we are chiefly interested in image segmentation applications, we focus on clustering methods that are monothetic, partitional, and hierarchical.  The term {\em hierarchical clustering\/} was first coined in~\citep{johnson67}.  A hierarchical clustering can be viewed as a sequence of nested clusterings such that a cluster at a given level is either identical to a cluster already existing at the previous level or is formed by unioning two or more clusters existing at the previous level.  It is convenient to represent this hierarchy in the form of a tree called {\em dendrogram\/}~\citep{sokal-sneath63} or {\em taxonomic tree\/} (see \citep{sneath57} for this latter terminology as well as a procedure which in essence already defined the concept of hierarchical clustering). The first detailed study about the use of trees in the context of hierarchical clustering appeared in \citep{hartigan67}.

\begin{figure}
\centerline{\includegraphics[width=4cm]{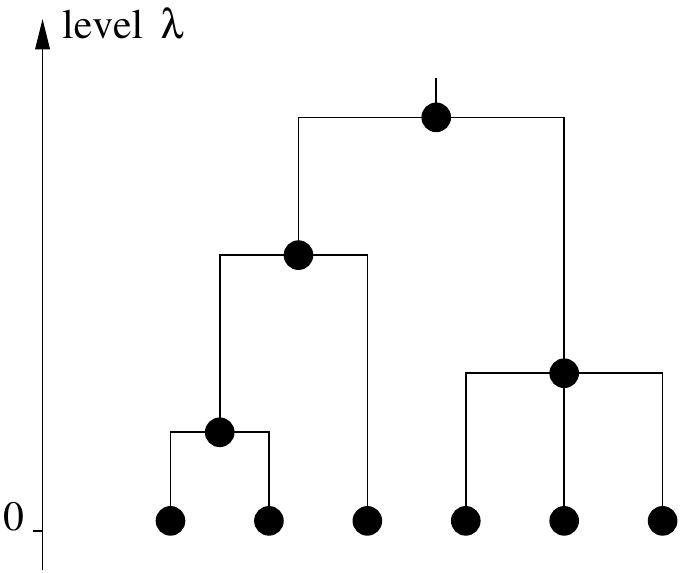}}
\caption{\label{f.dendro}
An example of dendrogram starting from 6 objects at the bottom of the hierarchy (level $\lambda=0$).  At the top of the hierarchy, there remains only one cluster containing all objects.}
\end{figure}

By construction, a hierarchical clustering is parameterised by a non-negative real number $\lambda$ indicating the level of a given clustering in the hierarchy.  At the bottom level, this number is equal to zero and each object correspond to a cluster so that the finest possible partition is obtained.  At the top level only one cluster containing all objects remains.  Given any two objects, it is possible to determine the minimum level value for which these two objects belong to the same cluster.  A key property of hierarchical clustering is that the function that measures this minimum level is an {\em ultrametric\/}.  An ultrametric is a measurement that satisfies all properties of a metric (distance) plus a condition stronger than the triangle inequality and called ultrametric inequality.  It states that the distance between two objects is lower than or equal to the maximum of the distances calculated from (i)~the first object to an arbitrary third object and (ii)~this third object to the second object.  Denoting by $\umetric$ the ultrametric function and $p$, $q$, and $r$ respectively the first, second and third objects, the ultrametric inequality corresponds to the following inequality:
\[
\umetric(p, q) \leq \max\{\umetric(p, r), \umetric(r, q)\}.
\]
The ultrametric property of hierarchical clustering was discovered simultaneously in \citep{johnson67,jardine-jardine-sibson67}, see also \citep{benzecri73} for a thorough study on ultrametrics in classification.  An example of dendrogram is displayed in Fig.~\ref{f.dendro}.

The measure of similarity between the input objects requires the
selection of a dissimilarity measurement.  A dissimilarity measurement
between the elements of a set $V$ is a function $\diss$ from $V\times
V$ to the set of nonnegative real numbers satisfying the three
following conditions: (i)~$\diss(p,q)\geq 0$ for all $p,q\in V$ (i.e.,
positiveness), (ii)~$\diss(p,p)=0$ for all $p\in V$, and
(iii)~$\diss(p,q)=\diss(q,p)$ for all $p,q\in V$ (i.e., symmetry).
Starting from an arbitrary dissimilarity measurement, it is possible
to construct a hierarchical clustering: if the dissimilarity is
increasing with the merging order, an ultrametric distance between any
two objects (or clusters) can be defined as the dissimilarity
threshold level from which these two objects (or clusters) belong to
the same cluster; if if the dissimilarity is {\em not\/} increasing
with the merging order, then any increasing function of the merging
order can be used.

In practice, the hierarchy is constructed by an iterative procedure merging first the object pair(s) with the smallest dissimilarity value so as to form the first non-trivial cluster(s) (i.e., non reduced to one object).  To proceed, the dissimilarity measurement between objects needs to be extended so as to be applicable to clusters.  Let $\acluster_i$ and $\acluster_j$ denote two clusters obtained at a given iteration level.  The dissimilarity between between these two clusters is naturally defined as a function of the dissimilarities between the objects belonging to these clusters:
\[
\diss(\acluster_i,\acluster_j)=f\{\diss(p,q)) \mid p\in\acluster_i \mbox{ and $q\in\acluster_j$}\}.
\]
Typical choices for the function $f$ are the minimum or maximum.  The maximum rule leads to the complete-linkage clustering (sometimes called maximum method) and dates back to~\citep{sorensen48}.  Complete-linkage is subject to ties in case the current smallest dissimilarity value is shared by two or more clusters. 
Consequently, one of the possible merge must be chosen and often this can only be achieved by resorting to some arbitrary (order dependent or random) selection.  By construction, complete-linkage favours compact clusters.
On the other hand, the minimum rule is not subject to ties (and is therefore uniquely defined) and does not favour compact clusters.  The resulting clustering is called the single-linkage clustering\footnote{The concept of single-linkage and its use for classification purposes were apparently suggested for the first time in \citep{florek-etal51} while the terminology single-linkage seems to be due to Sneath, see~\cite[p.~180]{sokal-sneath63} where it is also called Sneath's method.}~(sometimes called minimum method).  Indeed, only the pair (link) with the smallest dissimilarity value is playing a role.  

The single-linkage clustering is closely related to the minimum
spanning tree~\citep{gower-ross69},
defined as follows. To any edge-weighted graph $X$, the number $F(X)=\sum_{u=\in
  E(X)}F(u)$ is the {\em weight} of the graph.  
A spanning tree of a connected graph $X$ is a graph whose vertex set is equal to $V(X)$ and whose edge set is a subset of  $E(X)$ such that no cycles are formed.  A spanning tree of $X$ with minimum weight is called a {\em minimum spanning tree\/} of $X$. 

Indeed, the hierarchy underlying the single-linkage clustering is at the root of the greedy algorithm of Kruskal \citep{kruskal56} for solving the minimum spanning tree problem\footnote{The first explicit formulation of the minimum spanning tree problem is attributed to \citep{boruvka26}, see detailed account on the history of the problem in~\citep{graham-hell85}.}.  In this algorithm, referred to as 'construction~A' in \citep{kruskal56}, the edges of the graph are initially sorted by increasing edge weights (in a clustering perspective, the nodes of the graph are the objects and the edge weights are defined by the dissimilarity measurements between the objects).  Then, a minimum spanning tree $MST$ is defined recursively as follows: the next edge is added to $MST$ if and only if together with $MST$ it does not form a circuit.  That is, there is a one-to-one correspondence between (i)~the clusters obtained for a given dissimilarity level and (ii)~the subtrees obtained for a distance equal to this level in Kruskal's greedy solution to the minimum spanning tree problem.

While the single-linkage is not subject to ties, it is sensitive to the presence of objects of intermediate characteristics (transitions) that may occur between two clearly defined populations, see~\citep{wishart69} for a detailed discussion as well as Sec.~\ref{s.transp}.
This effect is sometimes called 'chaining-effect' although this latter terminology is somewhat misleading for chaining is the very principle of single-linkage~\citep{jardine-sibson68h-nh}.

\section{Hierarchical image segmentation}\label{s.his}
After a brief discussion on the definition of image segmentation and hierarchical image segmentation (see Sec.~\ref{ss.ishis}), methods relying on graph representations are presented (Sec.~\ref{ss.hisbgraph}) and then those developed in MM (Sec.~\ref{ss.hisbgraph}).

\subsection{From image segmentation to hierarchical image segmentation}\label{ss.ishis}
A segmentation of the definition domain $V$ of an image is usually
defined as a partition of $V$ into disjoint connected subsets
$V_i,\ldots,V_n$ (called segments) such that there exists a logical
predicate $P$ returning true on each segment but false on any union of
adjacent segments~\citep{horowitz-pavlidis74,zucker76}.  That is, a
series of subsets $V_i$ of the definition domain $V$ of an image forms
a segmentation of this image if and only if the following four
conditions are met (i)~$\cup_i(V_i)=X$, (ii)~$V_i\cap V_j=\emptyset$
for all $i\neq j$, (iii)~$P(V_i)=\mathrm{true}$ for all $i$, and
(iv)~$P(V_i\cup V_j)=\mathrm{false}$ if $V_i$ and $V_j$ are adjacent.
The first condition requires that every picture element (pixel) must
belong to a segment. The second condition requires that each segment
does not overlap any other segment.  The third condition determines
what kind of properties each segment must satisfy, i.e., what
properties the image pixels must satisfy to be in the same segment.
The fourth condition ensures that the segments are maximal in the
sense that specifies that any merging of any adjacent regions would
violate the third condition.

Note that uniqueness of the resulting segmentation given a predicate is not required.  If uniqueness is desired, the predicate should rely on an equivalence relation owing to the one-to-one correspondence between the unique partitions of a set and the equivalence relations on it, see for example \cite[p.~48]{jardine-sibson71}.  Interestingly,  the relation 'is connected' is an equivalence relation since it is reflexive (a point is connected to itself by a path of length $0$), symmetric (if a point $p$ is connected to a point $q$ then $q$ is connected to $p$ since the reversal of a path is path), and transitive (if $p$ is connected to $q$ and $q$ to $r$ then $p$ is connected to $r$ since the concatenation of two paths is a path).  Any given connectivity relation partitions the set of pixels of a given input image into equivalent classes called connected components \citep{rosenfeld79}.  They are maximal subsets of pixels such that every pair of pixels belonging to such a subset is connected.  The resulting partition meets therefore all conditions of a segmentation.

The segments resulting from a segmentation procedure are analogous to the clusters obtained when clustering data.  Clustering techniques can be applied to image data for either classification or segmentation purposes.  In the former case, the spatial position of the pixels does not necessarily play a role for clusters are searched in a parametric space such as the multivariate histogram.  The resulting clusters partition the parametric space into a series of classes and this partition is used as a look-up-table to indicate the class of each pixel of the input image.  An example of this approach using morphological clustering is proposed in~\citep{soille96jei}.  Contrary to data clustering applied to non-spatial data, the dissimilarity measurements between the data samples (i.e., the pixels) are not measured between all possible pairs.  Indeed, the spatial position of the pixels plays a key role so that measurements are only performed between adjacent pairs of pixels.  That is, the full dissimilarity matrix is very sparse: for a image of $m\times n$ pixels, there are $2mn-m-n$ entries in the $(mn)^2\times (mn)^2$ dissimilarity matrix when considering 4-adjacency relation.

By analogy with hierarchical clustering, hierarchical segmentation can be defined as a family of fine to coarse image partitions (i.e., family of ordered partitions) parameterised by a non-negative real number indicating the level of a given partition in the hierarchy.  Hierarchical segmentation is useful to help the detection of objects in an image.  In particular, it can be used to simplify the image in such a way that the elementary picture elements are not anymore the pixels but connected sets of pixels.  Indeed, in image data, analogues to phonemes and characters correspond to structural primitives that compress the data to a manageable size without eliminating any possible final interpretations~\citep{ahuja95}.  It should be emphasised that a hierarchical segmentation does not necessarily deliver segments directly corresponding to the searched objects.  This happens for instance when an object is not characterised by some homogeneity/separation criteria but from the consideration of an a priori model of the whole object (e.g. perceptual grouping and Gestalt theory).

There exists a fundamental difference between segmentation and classification.  Indeed, contrary to classification, segmentation requires the explicit definition of an adjacency graph or, more generally, a connection~\citep{Serra-2006,Ronse-2008}.  Typically, the $k$-nearest neighbouring graph with $k$ equal to 4 or 8 is used for processing 2-dimensional images.  With classification, a decision about the class (i.e., label) of each pixel can be reached without using its spatial context (position) so that it does not necessarily need the definition of an adjacency graph.  Nevertheless, any classification can be used to generate a segmentation.  Indeed, once an adjacency graph is added to the classified image, the maximal connected regions of pixels belonging to the same class generate a segmentation of the image definition domain.  If the considered adjacency graph is the complete graph, a one-to-one correspondence between the classes and the resulting connected components is obtained.

Hereafter, a selection of techniques achieving hierarchical image segmentation is proposed, extending the initial survey proposed in~\citep{soille2008pami}.  We start with generic methods based on graph representations and then proceed with specific methods developed in the context of mathematical morphology.  Recent developments related to constrained connectivity and ultrametric watersheds are discussed in Sec.~\ref{s.ccuw}.

\subsection{Methods based on graph representations}\label{ss.hisbgraph}
Horowitz and Pavlidis \citep{horowitz-pavlidis74,horowitz-pavlidis76} are among the first to suggest a formulation of hierarchical image segmentation in a graph theoretical framework.  It is based on the split-and-merge algorithm.  Because their implementation relies on a regular pyramid data structure with square blocks, it is not translation invariant and it favours blocky edges owing to the initial regular split of the image. In addition, the grouping stage of split-and-merge algorithms is order dependent, a drawback of all procedures updating the features of a region once new points are added to it. 

The idea of applying the single-linkage clustering method to produce hierarchical image segmentation was implemented for the first time by Nagao \citep{nagao-matsuyama-ikeda79,nagao-matsuyama80} for processing aerial images using grey level differences between adjacent pixels as dissimilarity measurement.  For colour images, the resulting dissimilarity vector led to the notion of differential threshold vector in~\citep{baraldi-parmiggiani96}.  The application of single-linkage clustering to image data are further developed in \citep{morris-etal86} using a graph theoretic framework.  This latter paper also details a minimax SST (Shortest Spanning Tree) segmentation allowing for the initial minimum spanning tree to be partitioned into n subtrees by recursively splitting the subtree with the larger cost into 2 subtrees (see also recursive SST segmentation into n regions).  Note that single-linkage clustering based on grey level difference dissimilarity was rediscovered much later in morphological image processing under the term quasi-flat zones \citep{meyer-maragos99,meyer-maragos2000}.  More recently, the more general and appropriate term of $\alpha$-connected component was proposed in~\citep{soille2008pami} to refer to any connected component (i.e., maximal set of connected pixel) of pixels such that any pair of pixels of this connected component can be linked by a path such that the dissimilarity value between two successive pixels of the path does not exceed a given dissimilarity threshold value (see details in Sec.~\ref{ss.conscon}).  The ultrametric behind the single-linkage hierarchical image segmentation is analogous to the one defined for single-linkage clustering, see Sec.~\ref{s.hclustering}.  

The hierarchy of graphs (irregular pyramids) proposed recently in \citep{nacken95,kropatsch-haxhimusa2004} builds on the graph weighted partitions developed in \citep{felzenszwalb-huttenlocher98,felzenszwalb-huttenlocher2004} and inspired by the seminal work of Zahn \citep{zahn71} on point data clustering and its extension to graph cut image segmentation in \citep{wu-leahy93,shi-malik2000}.  It relies on weighted graphs where each element of the edge set is given a weight corresponding to the range of the values of its two nodes.  The internal contrast of a connected component corresponds to the largest weight of all edges belonging to this connected component (an edge belongs to a connected component if its corresponding nodes belong to it or, alternatively, to a spanning tree of minimum sum of edge weights).  The external contrast is defined as the smallest weight of the edges linking a pixel of the considered connected component to another one.  The hierarchy is achieved by defining a dissimilarity measure accounting for both the internal and external contrasts.  The successive levels of the hierarchy are then obtained by iteratively merging the adjacent connected components of minimum dissimilarity.
An up-to-date survey (including comparisons) of both regular and irregular pyramidal structures can be found in \citep{marfil-etal2006}.  A survey on graph pyramids for hierarchical segmentation is proposed in~\citep{kropatsch-haxhimusa-ion2007}.

The hierarchical image segmentation based on the notion of the cocoons of a graph relies on a complete-linkage hierarchy and its corresponding ultrametric~\citep{guigues-lemen-cocquerez2003}.  The same authors introduced the notion of scale-sets~\citep{guigues-cocquerez-lemen2006} where the dissimilarity measurement is replaced by a two-term energy minimization process where the first term accounts for the amount of information required to encode the deviation of the data against the region model (typically taken as the mean of the region) and the second term is proportional to the amount of information required to encode the shape of the model (typically taken as the boundary length of the region).  

In \citep{arbelaez-cohen2004}, the extrema mosaic (influence zones of the image regional extrema) is considered as the base level of the hierarchy.  The dissimilarity between the segments is defined as the average gray level difference along the common boundary of these segments.  This dissimilarity is increasing with the merging order and is therefore an ultrametric.  Generic ultrametric distances obtained by integrating local contour cues along the regions boundaries and combining this information with region attributes are proposed in~\citep{arbelaez2006ultrametric}.

\subsection{Methods developed in mathematical morphology}\label{ss.hisbmm}
Mathametical morphology relies on the notion of lattices, and a theory
devoted to segmentation in this context recently appears~\cite{Serra-2006,Ronse-2008}.
From a practical point of view, most of the application schemes use
either a watershed-based approach or a tree-based approach.

\subsubsection{Watershed based}
The waterfall algorithm~\citep{beucher90phd,beucher94ismm,beucher-meyer93} can be considered as the first morphological hierarchical image segmentation method.  The elementary components of the base level of the tree underlying the waterfall hierarchy are the catchment basins of the gradient of the image.  Each basin is then set to the height of the lowest watershed pixel surrounding this basin while the watershed pixels keep their original value.  The watersheds of the resulting image delivers basins corresponding to the subsequent level of the hierarchy.  The procedure is then iterated until only one basin matching the image domain is obtained.  This hierarchy of partitions can be implemented directly on graph data structures as detailed in~\citep{vincent-soille91}.

Watershed hierarchies using the notion of contour dynamic is proposed in~\citep{najman-schmitt96}.  The arcs of the watersheds of the gradient of the original image are valued by their contour dynamic.  More precisely, the contour dynamic of an arc of a watershed separating two basins is defined as the height difference between the lowest point of this arc and the height of the highest regional minimum associated with these two basins.  The contour dynamic is a dissimilarity that satisfies all properties of an ultrametric.  The resulting contour dynamic map is a saliency map representing a hierarchy.  Indeed, a fine to coarse family of partitions is obtained by thresholding the contour dynamic map for increasing contour dynamic values.  By associating other dissimilarity measures to the arcs of the watersheds, other partition hierarchies are obtained. 

Note that, if one wants to obtain theoretical results
associating definitions and properties~\cite{CouNaj11}, one has to
work on edge-weighted graphs with the watershed-cut definition
\cite{cousty-etal2010pami} that links the watershed with the minimum spanning
tree as initially pointed out in~\cite{meyer94ismm}.

\subsubsection{Tree based}

Another type of hierarchy is obtained by considering the flat zones of
the image as the finest partition and then iteratively merging the
most similar flat zones.  This resulting tree is called binary
partition trees in~\citep{salembier-garrido2000}.  The tree always
represents a hierarchy indexed by the merging order and not always the
dissimilarity since the one used in~\citep{salembier-garrido2000} is
not an ultrametric.

Another tree, known as the component tree~\citep{jones97nsip,jones99} of the vertices (called max-tree or min-tree in~\citep{salembier-oliveras-garrido98} depending on whether its leaves are matching the image maxima or minima) represents the hierarchy of the level sets of the image and are therefore not directly representing a hierarchy of partitions of the image definition domain.

However, when defined not on the vertices but on the edges, we will
see below that the component tree is indeed a dendrogram representing
a hierarchy of connected partitions.

Reviews on hierarchical methods developed in mathematical morphology
based on watersheds are presented
in~\citep{meyer2001ijpr,IGMI_MeyNaj10}, and on trees
in~\cite{SalWilk-2009,Sal-2010}.  Recent developments related to
constrained connectivity and ultrametric watersheds are developed in
the next section.

\section{Constrained connectivity and ultrametric watersheds}\label{s.ccuw}

\subsection{Constrained connectivity}\label{ss.conscon}
\subsubsection{Preliminaries}
Let us first recall the notion of $\arangepsl$-connectivity that corresponds to single-linkage clustering applied to image data, see Sec.~\ref{ss.hisbgraph}.  Two pixels $p$ and $q$ of an image $f$ are $\arangepsl$-connected if there exists a path going from $p$ to $q$ such that the dissimilarity between any two successive pixels of this path does not exceed the value of the local parameter $\arangepsl$.  By definition, a pixel is $\arangepsl$-connected to itself.  Accordingly, the $\arangepsl$-connected component of a pixel $p$ is defined as the set of image pixels that are $\arangepsl$-connected to this pixel.  We denote this connected component by $\arangepsl\RCC(p)$:
$\arangepsl\RCC(p)=\{p\}\cup\big\{q\;|$ there exists a path $\apath=\langle p=p_1,\ldots,p_n=q\rangle$, $n>1$, such that $F(\{p_i,p_{i+1}\})\leq \arangel$ for all $1\leq i< n\big\}$.
In the case of grey level images and when considering the absolute intensity difference as dissimilarity measure, the $\alpha$-connected components of an image are equivalent to its quasi-flat zones~\citep{meyer-maragos99,meyer-maragos2000}.
Note that the edges of the connected graph corresponding to a given $\alpha$-connected component is defined by the pairs of adjacent pixels belonging to this $\alpha$-connected component such that their associated dissimilarity (weight) does not exceed $\alpha$.

\subsubsection{Definitions and properties}
The constrained connectivity paradigm~\citep{soille2007iciap,soille2008pami} originated from the need to develop a method preventing the formation of $\arangepsl$-connected components whose range values exceed that specified by the local range parameter $\arangepsl$ (assuming that the dissimilarity between two pixels is the absolute difference of their intensity values, see \citep{soille2011ismm,gueguen-soille2011ismm} for other examples of dissimilarity measures).  This is simply achieved by looking for the largest $\arangepsl$-connected components satisfying a global range constraint referred to as the global range parameter denoted by $\arangepsg$:
\[
\locglops\RCC(\apixp)=\bigvee\Big\{\arangel_i\RCC(\apixp)\;\Big|\;\arangel_i\leq\arangepsl
\mbox{ and $\range\Big(\arangel_i\RCC(\apixp)\Big)\leq\arangepsg$}\Big\},
\]
where the range function $\range$ calculates the difference between the maximum and the minimum values of a nonempty set of intensity values.  Note that the $\locglops$-connected components for $\arangepsl\geq\arangepsg$ are equivalent to those obtained for $\arangepsl=\arangepsg$.  That is, when $\arangepsl\geq\arangepsg$ the local range parameter does not play a role.  This leads to the concept of $(\arangepsg)$-connected component\footnote{The parenthesis is not dropped to avoid confusion with $\arangepsl$-connected components when the Greek letters are replaced by a numerical value indicating the actual value of the corresponding range parameter.}:
\[
(\arangepsg)\RCC(p)=(\arangepsl\geq\arangepsg,\arangepsg)\RCC(p)=
\bigvee\Big\{\arangepsl_i\RCC(p)\;|\;\range\Big(\arangel_i\RCC(\apixp)\Big)\leq\arangepsg\Big\}.
\]
The corresponding global dissimilarity measurement $\dissglo$ between two pixels is defined by the smallest range of the $\arangel$-connected components containing these two pixels.  This dissimilarity measurement satisfies also the ultrametric inequality.  Accordingly, we obtain the following equivalent definition of a $(\arangepsg)$-connected component: $(\arangepsg)\RCC(p)=\{q\;|\;\dissglo(p,q)\leq \arangepsg\}$.  In contrast to what happens with the local dissimilarity measurement $\dissloc$, the range of the values of arbitrary pairs of pixels belonging to the same $(\arangepsg)$-connected component is limited, the maximal value of this range being equal to $\arangepsg$.  Therefore, the resulting clustering bears some resemblance to the complete linkage clustering suggested in \citep{sorensen48} but, contrary to the latter procedure, it is unequivocal (see \cite[pp.~181-182]{sokal-sneath63} for an account on the equivocality of the complete linkage clustering).  The generalisation of the concept of constrained connectivity to arbitrary constraints is presented in~\citep{soille2007iciap}.

\subsubsection{Separation value}
The separation value $\separmin$ of an iso-intensity connected component (flat-zone) can be defined in terms of grey tone hit-or-miss transforms~\citep{soille2002dgci} with {\em adaptive\/} composite structuring elements.  The adaptive hit-or-miss transform of a pixel with the composite structuring element containing the origin $\origin$ for the foreground component and its direct neighbours having a strictly lower value $\ngbl(\origin)$ for the background component outputs the difference between the input pixel value and that of its largest lower neighbour(s) if the set of its lower neighbours is non-empty, 0 otherwise.  This adaptive hit-or-miss transform is denoted by $\hmtop_{(\origin,\ngbl(\origin))}$:
\[
[\hmtop_{(\origin,\ngbl(\origin))}](\apixp)=
\left\{
\begin{array}{ll}
f(\apixp)-\vee\{f(\apixq)\mid\apixq\in\ngbl(\apixp)\}&\mbox{if $\ngbl(\apixp)\neq\emptyset$,}\\
0& \mbox{otherwise}.
\end{array}
\right.
\]
Similarly, the adaptive hit-or-miss transform $\hmtop_{(\ngbg(\origin),\origin)}$ of a pixel outputs the difference between the value of its smallest greater neighbour(s) and that of the pixel itself, if the set of its greater neighbours  $\ngbg(\origin)$ is non-empty, 0 otherwise:
\[
[\hmtop_{(\ngbg(\origin),\origin)}](\apixp)=
\left\{
\begin{array}{ll}
\wedge\{f(\apixq)\mid\apixq\in\ngbg(\apixp)\}-f(\apixp)&\mbox{if $\ngbg(\apixp)\neq\emptyset$},\\
0& \mbox{otherwise}.
\end{array}
\right.
\]
The non-zero values of the point-wise minimum between the two hit-or-miss transforms corresponds to the transition pixels in the sense that these pixels have simultaneously lower and greater neighbours (and the point-wise minimum image indicates the minimum height of the transition). The binary mask of transition pixels can therefore be obtained by the following operator denoted by $\transiop$:
\[
\transiop=
\thresh_{> 0}[\hmtop_{(\origin,\ngbl(\origin))}\wedge\hmtop_{(\ngbg(\origin),\origin)}].
\]
In \citep{soille-grazzini2009ismm}, the same mask is obtained by considering the non-zero values of the point-wise minimum of the gradients by erosion and dilation with the elementary neighbourhood (the pixel and its direct neighbours) as structuring element.  In this latter case, the point-wise minimum image indicates the maximum height of the transition.

The minimum separation value of a pixel of an image is defined as the minimum intensity difference between a pixel and its neighbour(s) having a different value from this pixel if such neighbour(s) exist, 0 otherwise.  It is denoted by $[\separmin(f)](\apixp)$ and can be calculated as follows:
\[
[\separmin(f)](\apixp)=
\left\{
\begin{array}{ll}
[\hmtop_{(\origin,\ngbl(\origin))}(f)](\apixp)&
\mbox{if $[\hmtop_{(\origin,\ngbl(\origin))}(f)](\apixp)<[\hmtop_{(\ngbg(\origin),\origin)}(f)](\apixp)$}\\ & \mbox{and $[\hmtop_{(\origin,\ngbl(\origin))}(f)](\apixp)\neq 0$,} \\
\mbox{$[\hmtop_{(\ngbg(\origin),\origin)}(f)](\apixp)$}& \mbox{otherwise}.
\end{array}
\right.
\]
The minimum separation value of an iso-intensity connected component $0\RCC$ is then defined as the smallest (minimum) separation value of its pixels:
\[
\separmin(0\RCC)=\wedge\{\separmin(\apixq)\mid \apixq\in0\RCC\mbox{ and $\separmin(\apixq)\neq 0$}\}.
\]
It is equivalent to the smallest $\arangepsl$ value such that $\arangepsl\RCC\neq 0\RCC$.  Similarly, the operator that sets each pixel of the image to the minimum separation value of the iso-intensity connected component it belongs to is defined as follow:
\[
[\separmin(0\RCC(f))](\apixp)=
\wedge\{\separmin(\apixq)\mid \apixq\in0\RCC(\apixp)\mbox{ and $\separmin(\apixq)\neq 0$}\}.
\]
It can be viewed as an adaptive operation where the output value at a given pixel depends on the iso-intensity component of this pixel and the neighbouring pixels of this component.  By replacing  the $\wedge$ operation with the $\vee$ operation in the minimum separation definitions, we obtain the definitions for maximal separations.  Figure~\ref{f.extisol} illustrates the map of minimal separation of the pixels and iso-intensity connected components of a synthetic image.
\begin{figure}
\hspace*{\fill}\setlength{\unitlength}{0.5cm}
\begin{picture}(7,7)(0,0)
\setcoordinatesystem units <0.5cm,0.5cm> point at 0 0
\setplotarea x from 0 to 7, y from 0 to 7
\grid {7} {7}
\put(0.5,0.5){\makebox(0,0)[c]{0}}
\put(1.5,0.5){\makebox(0,0)[c]{2}}
\put(2.5,0.5){\makebox(0,0)[c]{9}}
\put(3.5,0.5){\makebox(0,0)[c]{3}}
\put(4.5,0.5){\makebox(0,0)[c]{8}}
\put(5.5,0.5){\makebox(0,0)[c]{5}}
\put(6.5,0.5){\makebox(0,0)[c]{9}}
\put(0.5,1.5){\makebox(0,0)[c]{1}}
\put(1.5,1.5){\makebox(0,0)[c]{0}}
\put(2.5,1.5){\makebox(0,0)[c]{8}}
\put(3.5,1.5){\makebox(0,0)[c]{4}}
\put(4.5,1.5){\makebox(0,0)[c]{9}}
\put(5.5,1.5){\makebox(0,0)[c]{6}}
\put(6.5,1.5){\makebox(0,0)[c]{7}}
\put(0.5,2.5){\makebox(0,0)[c]{3}}
\put(1.5,2.5){\makebox(0,0)[c]{2}}
\put(2.5,2.5){\makebox(0,0)[c]{7}}
\put(3.5,2.5){\makebox(0,0)[c]{9}}
\put(4.5,2.5){\makebox(0,0)[c]{9}}
\put(5.5,2.5){\makebox(0,0)[c]{1}}
\put(6.5,2.5){\makebox(0,0)[c]{1}}
\put(0.5,3.5){\makebox(0,0)[c]{1}}
\put(1.5,3.5){\makebox(0,0)[c]{1}}
\put(2.5,3.5){\makebox(0,0)[c]{9}}
\put(3.5,3.5){\makebox(0,0)[c]{3}}
\put(4.5,3.5){\makebox(0,0)[c]{4}}
\put(5.5,3.5){\makebox(0,0)[c]{2}}
\put(6.5,3.5){\makebox(0,0)[c]{6}}
\put(0.5,4.5){\makebox(0,0)[c]{1}}
\put(1.5,4.5){\makebox(0,0)[c]{0}}
\put(2.5,4.5){\makebox(0,0)[c]{4}}
\put(3.5,4.5){\makebox(0,0)[c]{1}}
\put(4.5,4.5){\makebox(0,0)[c]{1}}
\put(5.5,4.5){\makebox(0,0)[c]{2}}
\put(6.5,4.5){\makebox(0,0)[c]{5}}
\put(0.5,5.5){\makebox(0,0)[c]{2}}
\put(1.5,5.5){\makebox(0,0)[c]{1}}
\put(2.5,5.5){\makebox(0,0)[c]{9}}
\put(3.5,5.5){\makebox(0,0)[c]{8}}
\put(4.5,5.5){\makebox(0,0)[c]{8}}
\put(5.5,5.5){\makebox(0,0)[c]{9}}
\put(6.5,5.5){\makebox(0,0)[c]{1}}
\put(0.5,6.5){\makebox(0,0)[c]{1}}
\put(1.5,6.5){\makebox(0,0)[c]{3}}
\put(2.5,6.5){\makebox(0,0)[c]{8}}
\put(3.5,6.5){\makebox(0,0)[c]{7}}
\put(4.5,6.5){\makebox(0,0)[c]{8}}
\put(5.5,6.5){\makebox(0,0)[c]{8}}
\put(6.5,6.5){\makebox(0,0)[c]{2}}
\end{picture}
\hspace*{\fill}
\setlength{\unitlength}{0.5cm}
\begin{picture}(7,7)(0,0)
\setcoordinatesystem units <0.5cm,0.5cm> point at 0 0
\setplotarea x from 0 to 7, y from 0 to 7
\grid {7} {7}
\put(0.5,0.5){\makebox(0,0)[c]{1}}
\put(1.5,0.5){\makebox(0,0)[c]{2}}
\put(2.5,0.5){\makebox(0,0)[c]{1}}
\put(3.5,0.5){\makebox(0,0)[c]{1}}
\put(4.5,0.5){\makebox(0,0)[c]{1}}
\put(5.5,0.5){\makebox(0,0)[c]{1}}
\put(6.5,0.5){\makebox(0,0)[c]{2}}
\put(0.5,1.5){\makebox(0,0)[c]{1}}
\put(1.5,1.5){\makebox(0,0)[c]{1}}
\put(2.5,1.5){\makebox(0,0)[c]{1}}
\put(3.5,1.5){\makebox(0,0)[c]{1}}
\put(4.5,1.5){\makebox(0,0)[c]{1}}
\put(5.5,1.5){\makebox(0,0)[c]{1}}
\put(6.5,1.5){\makebox(0,0)[c]{1}}
\put(0.5,2.5){\makebox(0,0)[c]{1}}
\put(1.5,2.5){\makebox(0,0)[c]{1}}
\put(2.5,2.5){\makebox(0,0)[c]{1}}
\put(3.5,2.5){\makebox(0,0)[c]{2}}
\put(4.5,2.5){\makebox(0,0)[c]{5}}
\put(5.5,2.5){\makebox(0,0)[c]{1}}
\put(6.5,2.5){\makebox(0,0)[c]{5}}
\put(0.5,3.5){\makebox(0,0)[c]{2}}
\put(1.5,3.5){\makebox(0,0)[c]{1}}
\put(2.5,3.5){\makebox(0,0)[c]{2}}
\put(3.5,3.5){\makebox(0,0)[c]{1}}
\put(4.5,3.5){\makebox(0,0)[c]{1}}
\put(5.5,3.5){\makebox(0,0)[c]{1}}
\put(6.5,3.5){\makebox(0,0)[c]{1}}
\put(0.5,4.5){\makebox(0,0)[c]{1}}
\put(1.5,4.5){\makebox(0,0)[c]{1}}
\put(2.5,4.5){\makebox(0,0)[c]{3}}
\put(3.5,4.5){\makebox(0,0)[c]{2}}
\put(4.5,4.5){\makebox(0,0)[c]{1}}
\put(5.5,4.5){\makebox(0,0)[c]{1}}
\put(6.5,4.5){\makebox(0,0)[c]{1}}
\put(0.5,5.5){\makebox(0,0)[c]{1}}
\put(1.5,5.5){\makebox(0,0)[c]{1}}
\put(2.5,5.5){\makebox(0,0)[c]{1}}
\put(3.5,5.5){\makebox(0,0)[c]{1}}
\put(4.5,5.5){\makebox(0,0)[c]{1}}
\put(5.5,5.5){\makebox(0,0)[c]{1}}
\put(6.5,5.5){\makebox(0,0)[c]{1}}
\put(0.5,6.5){\makebox(0,0)[c]{1}}
\put(1.5,6.5){\makebox(0,0)[c]{2}}
\put(2.5,6.5){\makebox(0,0)[c]{1}}
\put(3.5,6.5){\makebox(0,0)[c]{1}}
\put(4.5,6.5){\makebox(0,0)[c]{1}}
\put(5.5,6.5){\makebox(0,0)[c]{1}}
\put(6.5,6.5){\makebox(0,0)[c]{1}}
\end{picture}
\hspace*{\fill}
\setlength{\unitlength}{0.5cm}
\begin{picture}(7,7)(0,0)
\setcoordinatesystem units <0.5cm,0.5cm> point at 0 0
\setplotarea x from 0 to 7, y from 0 to 7
\grid {7} {7}
\put(0.5,0.5){\makebox(0,0)[c]{1}}
\put(1.5,0.5){\makebox(0,0)[c]{2}}
\put(2.5,0.5){\makebox(0,0)[c]{1}}
\put(3.5,0.5){\makebox(0,0)[c]{1}}
\put(4.5,0.5){\makebox(0,0)[c]{1}}
\put(5.5,0.5){\makebox(0,0)[c]{1}}
\put(6.5,0.5){\makebox(0,0)[c]{2}}
\put(0.5,1.5){\makebox(0,0)[c]{1}}
\put(1.5,1.5){\makebox(0,0)[c]{1}}
\put(2.5,1.5){\makebox(0,0)[c]{1}}
\put(3.5,1.5){\makebox(0,0)[c]{1}}
\put(4.5,1.5){\makebox(0,0)[c]{1}}
\put(5.5,1.5){\makebox(0,0)[c]{1}}
\put(6.5,1.5){\makebox(0,0)[c]{1}}
\put(0.5,2.5){\makebox(0,0)[c]{1}}
\put(1.5,2.5){\makebox(0,0)[c]{1}}
\put(2.5,2.5){\makebox(0,0)[c]{1}}
\put(3.5,2.5){\makebox(0,0)[c]{1}}
\put(4.5,2.5){\makebox(0,0)[c]{1}}
\put(5.5,2.5){\makebox(0,0)[c]{1}}
\put(6.5,2.5){\makebox(0,0)[c]{1}}
\put(0.5,3.5){\makebox(0,0)[c]{1}}
\put(1.5,3.5){\makebox(0,0)[c]{1}}
\put(2.5,3.5){\makebox(0,0)[c]{2}}
\put(3.5,3.5){\makebox(0,0)[c]{1}}
\put(4.5,3.5){\makebox(0,0)[c]{1}}
\put(5.5,3.5){\makebox(0,0)[c]{1}}
\put(6.5,3.5){\makebox(0,0)[c]{1}}
\put(0.5,4.5){\makebox(0,0)[c]{1}}
\put(1.5,4.5){\makebox(0,0)[c]{1}}
\put(2.5,4.5){\makebox(0,0)[c]{3}}
\put(3.5,4.5){\makebox(0,0)[c]{1}}
\put(4.5,4.5){\makebox(0,0)[c]{1}}
\put(5.5,4.5){\makebox(0,0)[c]{1}}
\put(6.5,4.5){\makebox(0,0)[c]{1}}
\put(0.5,5.5){\makebox(0,0)[c]{1}}
\put(1.5,5.5){\makebox(0,0)[c]{1}}
\put(2.5,5.5){\makebox(0,0)[c]{1}}
\put(3.5,5.5){\makebox(0,0)[c]{1}}
\put(4.5,5.5){\makebox(0,0)[c]{1}}
\put(5.5,5.5){\makebox(0,0)[c]{1}}
\put(6.5,5.5){\makebox(0,0)[c]{1}}
\put(0.5,6.5){\makebox(0,0)[c]{1}}
\put(1.5,6.5){\makebox(0,0)[c]{2}}
\put(2.5,6.5){\makebox(0,0)[c]{1}}
\put(3.5,6.5){\makebox(0,0)[c]{1}}
\put(4.5,6.5){\makebox(0,0)[c]{1}}
\put(5.5,6.5){\makebox(0,0)[c]{1}}
\put(6.5,6.5){\makebox(0,0)[c]{1}}
\end{picture}
\hspace*{\fill}
\caption{\label{f.extisol}
Left: a synthetic  $7\times 7$ image $f$ with its intensity values~\cite[Fig.~2a]{soille2008pami}. Middle: the map of separation value of its pixels $\separmin(f)$. Right: the map of separation value of its flat zones $\separmin(0\RCC(f))$.
}
\end{figure}

The regional maxima $\rmax$ of $\separmin(0\RCC(f))$ can be used to flag the flat zones that are the most isolated.  Conversely, the regional minima $\rmin$ of $\separmin(0\RCC(f))$ can be used to flag the flat zones from which an immersion simulation should be iniated to compute the successive levels of the hierarchy of constrained components.  By doing so, an algorithm similar to the watershed by flooding simulation~\citep{soille-vincent90} can be designed.

\subsubsection{Alpha-tree representation}
Constrained connectivity relies on the definition of $\alpha$-connectivity.  The later boils down to the single-linkage clustering of the image pixels given the underlying dissimilarity measure between adjacent pixel pairs.  The corresponding single-linkage dendrogram was described as a {\em spatially rooted tree\/} in~\cite{soille2008pami}.  This spatially rooted tree was introduced as the {\em alpha-tree\/} in \cite{ouzounis-soille2011ismm,ouzounis-soille2011alphatree}.  It represents the fine to coarse hierarchy of partitions for an increasing value of the dissimilarity threshold $\alpha$.  The alpha-tree can also be seen as a component tree representing the ordering relations of the $\alpha$-connected components of the image.  The representation in terms of min-tree is developed in Sec.~\ref{ss.uw}.

In the case of constrained connectivity, a given $(\alpha,\omega)$-partition corresponds to the highest cut of the alpha-tree such that all the nodes below this cut satisfy the $\alpha$ and $\omega$ constraints.  Usually this cut is not horizontal.  A given $(\omega)$-partition corresponds to the highest cut of the alpha-tree such all the nodes below the cut satisfy the $\omega$ constraint.  Alternatively, a  $(\omega)$-partition can be obtained by performing a horizontal cut in the dendrogram based on the ultrametric $\dissglo$ (i.e., the omega-tree).  An example of omega-tree is given \cite{soille2010ijrs}.  Note however that the set of all $(\alpha,\omega)$-partitions is itself not ordered given the absence of order between arbitrary pairs of local and global dissimilarity threshold values.

\subsubsection{Edge-weighted graph setting and minimum spanning tree}\label{ss.ewgmst}
By construction, the connected components of the graph $\ewg{\alpha}=(V,\{\{p,q\}\in E\st F(\{p,q\})\leq\alpha\})$ are equivalent to the $\alpha$-connected components of $f$.  Since $\alpha$-connectivity corresponds to single-linkage clustering, there is an underlying minimum spanning tree associated to it (see also section~\ref{s.hclustering} and \citep{morris-etal86} for equivalent image segmentations based on the direct computation of a minimum spanning tree).  More precisely, the minimum spanning tree of the edge-weighted graph of an image is a tree spanning its pixels and such that the sum of the weights associated with the edges of the tree is minimal.  Denoting by $E_{\min}$ the edge set of a minimum spanning tree of the edge-weighted graph of an image, the connected components of the graph $(V,\{\{p,q\}\in E_{\min}\st F(\{p,q\})\leq\alpha\})$ are equivalent to those of $\ewg{\alpha}$ (equivalent in the sense that given any node, the set of nodes of the connected component of $(V,\{\{p,q\}\in E_{\min}\st F(\{p,q\})\leq\alpha\})$ containing this node is identical to the set of nodes of the connected component of $\ewg{\alpha}$ containing this very node).
Since the minimum spanning tree representation contains less edges than the initial edge-weighted graph, it is less memory demanding for further computations such as global range computations.  However, not all computations can be done on the minimum spannning tree (for example, connectivity constraints relying on the computation of a connectivity index \citep{soille2008pami} cannot be derived from it).

\subsection{Ultrametric watersheds: from hierarchical segmentations to saliency maps}\label{ss.uw}
We have several different ways to deal with hierarchies: dendrograms
and minimum spanning trees. In the case where a hierarchy is made of
connected regions, then we can also use its connected component tree, e.g., min-tree, max-tree or alpha-tree.  None of
these three tools allows for an easy visualisation of a given hierarchy
as an image.  We now introduce ultrametric
watershed~\citep{najman09:_ultram_water,IGMI_Naj11} as a tool
that helps visualising a hierarchy: we stack the contours of the
regions of the hierarchy; thus, the more a contour of a region is
present in the hierarchy, the more visible it is.  Ultrametric
watershed is the formalisation and the caracterisation of a notion
introduced under the name of {\em saliency
  map}~\citep{najman-schmitt96}.

\subsubsection{Ultrametric watersheds}

The formal definition of ultrametric watershed relies on the
topological watershed framework~\citep{bertrand2005}.

Let $X$ be a graph. An edge $u\in \overline{E(X)}$ is said to be {\em
  W-simple (for $X$)} if $X$ has the same number of connected
components as $X+u=(V(X),E(X)\cup\{u\})$.  

An edge $u$ such that $F(u)=\lambda$ is said to be {\em W-destructible
  (for $F$) with lowest value $\lambda_0$} if there exists $\lambda_0$
such that, for all $\lambda_1$, $\lambda_0< \lambda_1\leq \lambda$,
$u$ is W-simple for $\ewg{\lambda_1}$ and if $u$ is not W-simple for
$\ewg{\lambda_0}$.

A {\em topological watershed (on $G$)} is a map that contains no
W-destructible edges.

An {\em ultrametric watershed} is a topological watershed $F$ such
that $F(v)=0$ for any $v$ belonging to a minimum of $F$.

There exists a bijection between ultrametric distances and hierarchies
of partitions~\citep{johnson67}; in other word, to any hierarchy of
partitions is associated an ultrametric, and conversely, any
ultrametric yields a hierarchy of partitions, see also
Sec.~\ref{s.hclustering}. Similarly, there exists a bijection between
the set of hierarchies of {\em connected partitions} and the set of
ultrametric
watersheds~\citep{najman09:_ultram_water,IGMI_Naj11}.
In~\cite{ISMM_CouNaj2011}, it is proposed a generic algorithm for
computing hierarchies and their associated ultrametric watershed.

\subsubsection{Usage: gradient and dissimilarity}
Constrained connectivity is a hierarchy of flat zones of $f$, in the
sense where the $0$-connected components of $f$ are the zones of $f$
where the intensity of $f$ does not change. In a continuous world,
such zones would be the ones where the gradient is null, {\em i.e.}
$\nabla f=0$. However, the space we are working with is discrete, and
a flat zone of $f$ can consist in a single point. In general, it is
not possible to compute a gradient on the points or on the edges such
that this gradient is null on the flat zones. To compute a gradient on
the edges such that the gradient is null on the flat zones, we need to
``double'' the graph, for example we can do that by doubling the
number of points of $V$ and adding one edge between each new point and
the old one.

More precisely, if we denote the points of $V$ by
$V=\{p_0,\ldots,p_n\}$, we set $V'=\{p'_0,\ldots,p'_n\}$ (with $V\cap
V'=\emptyset$), and $E'=\{\{p_i,p'_i\}\st 0\leq i\leq n\}$. We then
set $V_1=V\cup V'$ and $E_1=E\cup E'$.

By construction, as $G=(V,E)$ is a connected graph, the graph
$G_1=(V_1,E_1)$ is a connected graph.  We also extend $f$ to $V'$, by
setting, for any $p'\in V'$, $f(p')=f(p)$, where $\{p,p'\}\in E'$.

We set, as in section~\ref{ss.conscon}, $F(\{p,q\})=|f(p)-f(q)|$.  The
map $F$ can be seen as the ``natural gradient'' of
$f$~\citep{Mattiussi-2000}. We can then apply the same scheme on this
$F$ as in section~\ref{ss.conscon} to find the hierarchy of $\alpha$-connected components.

We denote by $L(G_1)$ the edge graph (also called line graph) of $G_1$.  That is, each vertex of  $L(G_1)$ represents an edge of  $G_1$ and two vertices of $L(G_1)$ are adjacent if and only if their corresponding edges in $G_1$ share a common endpoint in $G_1$.  While the edges of $L(G_1)$ are not weighted, the weights of its nodes are given by the weights of the corresponding edges of $G_1$.  It follows that the minima of $L(G_1)$ are equivalent to the $0$-connected components of $G_1$.  More generally, the alpha-tree of $G_1$ is contained in the min-tree of $L(G_1)$.  Interestingly, the min-tree of $L(G_1)$ can be computed efficiently thanks to the quasi-linear algorithm described in~\cite{NajCou2006}.  Hence, the morphological framework of attribute filtering~\citep{breen-jones96a} can be applied to this min-tree~\citep{jones97nsip,salembier-oliveras-garrido98,jones99}, similarly to the segmentation of an image into $k$ regions proposed in~\citep{cousty-etal2009pami}.  This is in particular useful when the filtering is performed before computing a watershed and this is illustrated in the next paragraph for the computation of a hierarchy based on constrained connectivity.

Finding the $\locglops\RCC$s can be done by
filtering the ultrametric watershed $W$ of $F$ with $\range$ that acts
as a flooding on the topological/ultrametric watershed $W$ of $F$, and
then finding a (topological) watershed of the filtered
image. Repeating these steps for a sequence of ordered $(\alpha,\omega)$ vectors, we build a constrained connectivity
hierarchy. In effect, we are viewing a hierarchy as an image
(edge-weighted graph) and transforming it into another
hierarchy/image.

Thus, classical tools from mathematical morphology can be applied to
constrain any hierarchy. Similar examples exist in the literature, for
example~\citep{guigues-cocquerez-lemen2006}, where the authors compute
what they called a non-horizontal cut in the hierarchy, in other
words, they compute a flooding on a watershed. In their framework, the
flooding is controlled by an energy.

The advantages of using an ultrametric watershed are numerous. Let us
mention the two following ones:
\begin{enumerate}
\item an ultrametric watershed is visible. A dendrogram or a component
  tree can be drawn, but less information is available from such a
  drawing, and visualising a MST is not really useful;
\item an ultrametric watershed allows the use any information in the
  contours between regions; such information is not available on the
  component tree, and is only partially available with a MST (which
  contains only the pass between regions).

\end{enumerate}
Let us note that those concepts are theoretically equivalent: even
their respective computational time is in practice nearly identical;
thus we can choose the one the most adapted to the desired usage.

\begin{figure*}[!t]
\noindent  \begin{tabular}{cc}
    \subfigure[Original image]{
      \includegraphics[width=.48\textwidth,trim = 12 38 150 88,clip]{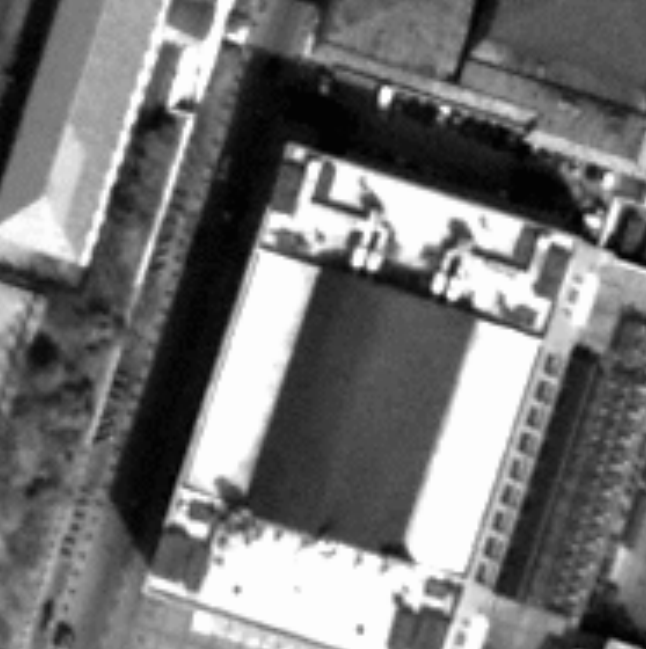}
    }
    &
    \subfigure[$W^1$(logarithmic grey-scale)]{
      \includegraphics[width=.48\textwidth,trim = 50 150 600 350,clip]{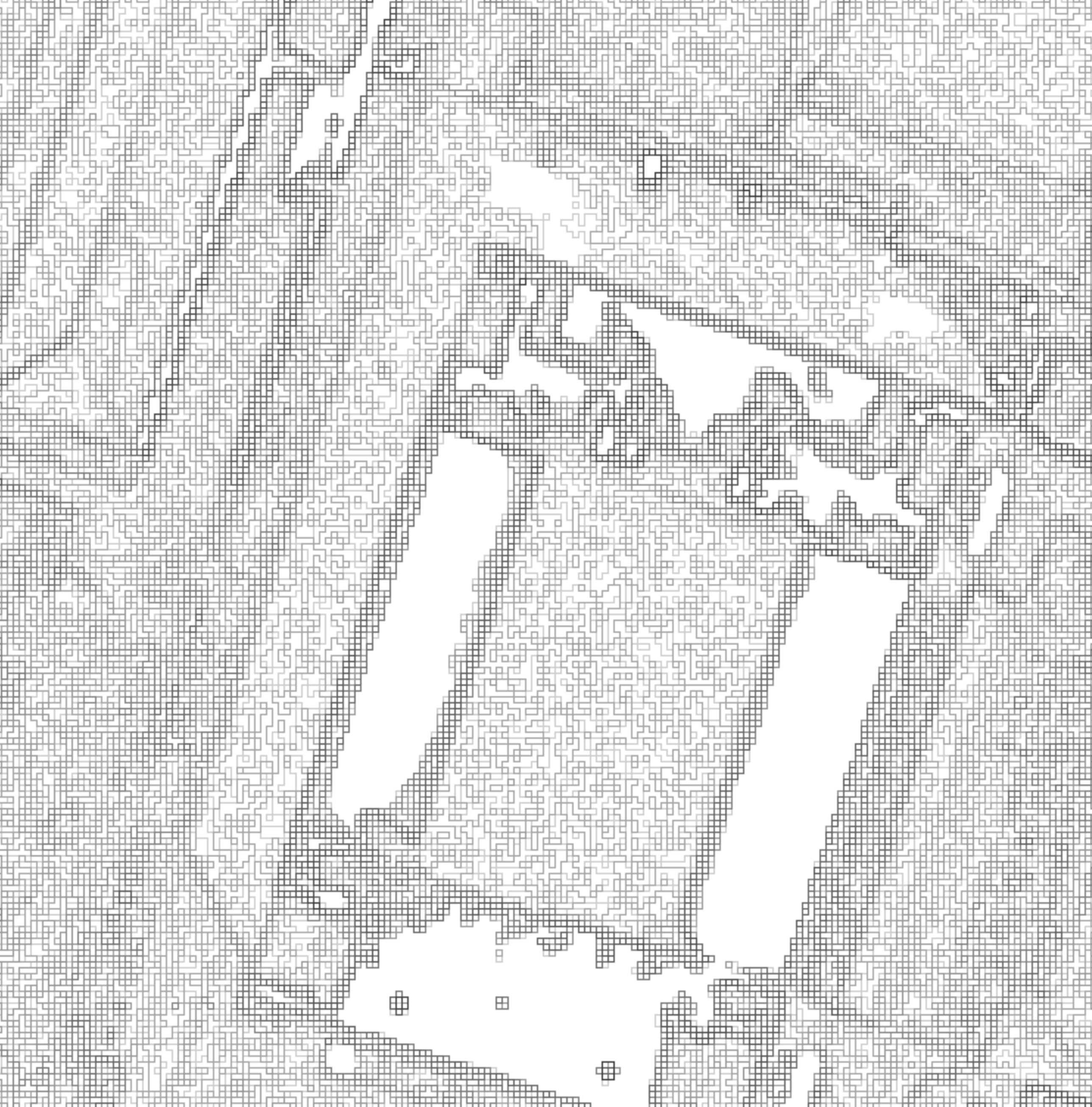}
    }
    \\
    \subfigure[$W^2$]{
      \includegraphics[width=.48\textwidth,trim = 50 150 600 350,clip]{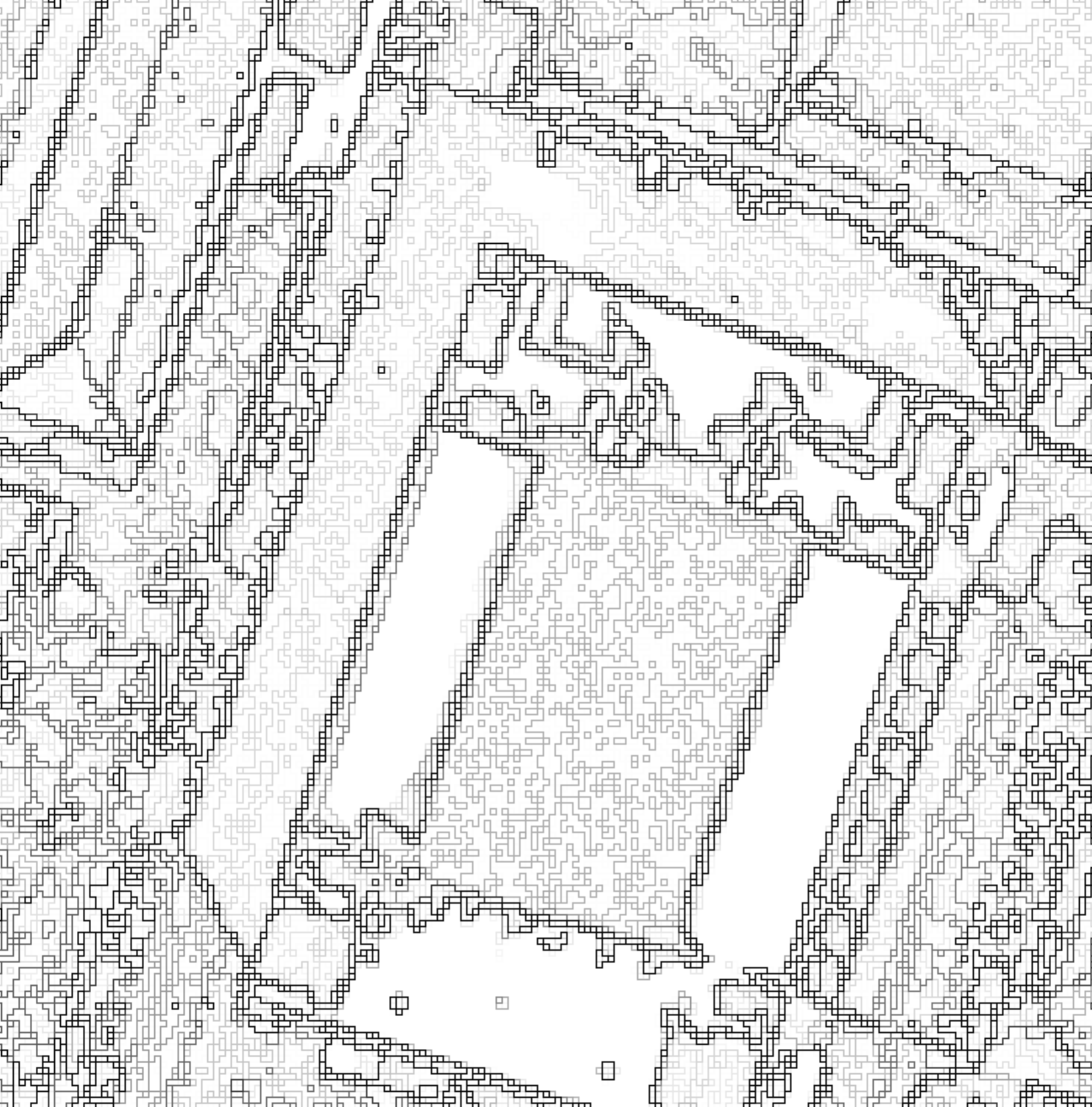}
    }
    &
    \subfigure[Area-filtering ultrametric watershed]{
      \includegraphics[width=.48\textwidth,trim = 50 150 600 350,clip]{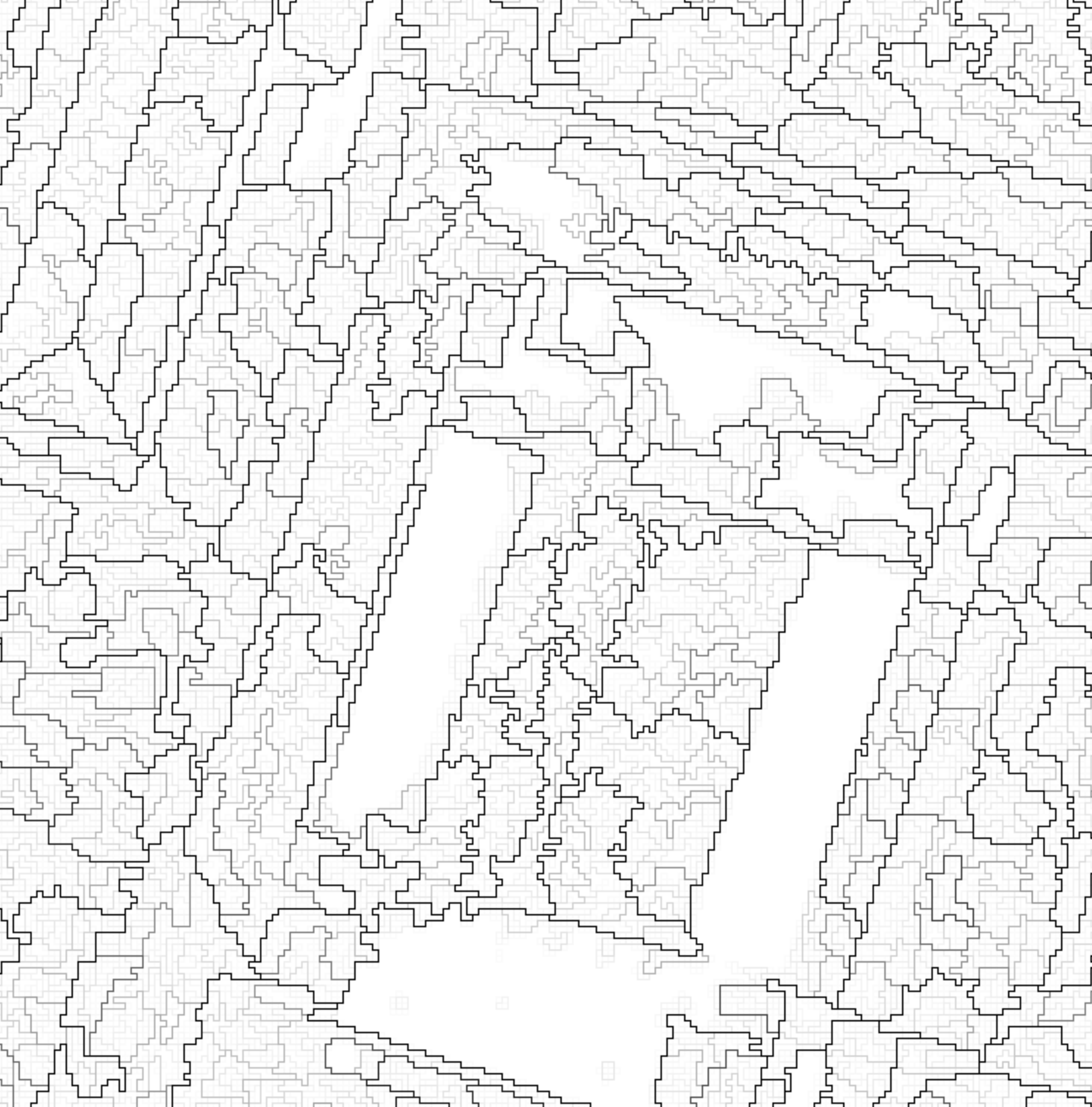}
    }
  \end{tabular}
 \caption{Constrained connectivity and ultrametric watersheds. (a) Original image (extract from the panchromatic channel of a Quickbird Imagery {\copyright} DigitalGlobe Inc., 2007, distributed by Eurimage). (b) Ultrametric watershed $W^1$ for  the $\alpha$-connectivity (the grey level of a contour corresponds to the $\alpha$ value above which the contour disappears in the $\alpha$-hierarchy. (c) Ultrametric watershed $W^2$ for the constrained connectivity (the grey level of a contour corresponds to the $\alpha=\omega$ value above which the contour disappears in the $(\alpha,\omega=\alpha)$-hierarchy). (d) Ultrametric watersheds corresponding to one of the possible hierarchies of area-filterings on $W^2$.}
 \label{figX:alphaomega}
\end{figure*}

Visualising the hierarchy of constrained connectivity as an
ultrametric watershed allows ones to assess some of its qualities. One can
notice in Fig.~\ref{figX:alphaomega}.c a large number of transition
regions (small undesirable regions that persist in the hierarchy),
which is the topic of the next section.

\section{Transition pixels\label{s.transp}}

\begin{figure}
\hspace*{\fill}\setlength{\unitlength}{0.5cm}
\begin{picture}(7,7)(0,0)
\setcoordinatesystem units <0.5cm,0.5cm> point at 0 0
\setplotarea x from 0 to 7, y from 0 to 7
\grid {7} {7}
\put(0.5,0.5){\makebox(0,0)[c]{0}}
\put(1.5,0.5){\makebox(0,0)[c]{1}}
\put(2.5,0.5){\makebox(0,0)[c]{0}}
\put(3.5,0.5){\makebox(0,0)[c]{1}}
\put(4.5,0.5){\makebox(0,0)[c]{8}}
\put(5.5,0.5){\makebox(0,0)[c]{7}}
\put(6.5,0.5){\makebox(0,0)[c]{8}}
\put(0.5,1.5){\makebox(0,0)[c]{1}}
\put(1.5,1.5){\makebox(0,0)[c]{0}}
\put(2.5,1.5){\makebox(0,0)[c]{1}}
\put(3.5,1.5){\makebox(0,0)[c]{2}}
\put(4.5,1.5){\makebox(0,0)[c]{7}}
\put(5.5,1.5){\makebox(0,0)[c]{8}}
\put(6.5,1.5){\makebox(0,0)[c]{7}}
\put(0.5,2.5){\makebox(0,0)[c]{0}}
\put(1.5,2.5){\makebox(0,0)[c]{1}}
\put(2.5,2.5){\makebox(0,0)[c]{0}}
\put(3.5,2.5){\makebox(0,0)[c]{3}}
\put(4.5,2.5){\makebox(0,0)[c]{8}}
\put(5.5,2.5){\makebox(0,0)[c]{7}}
\put(6.5,2.5){\makebox(0,0)[c]{8}}
\put(0.5,3.5){\makebox(0,0)[c]{1}}
\put(1.5,3.5){\makebox(0,0)[c]{0}}
\put(2.5,3.5){\makebox(0,0)[c]{1}}
\put(3.5,3.5){\makebox(0,0)[c]{4}}
\put(4.5,3.5){\makebox(0,0)[c]{7}}
\put(5.5,3.5){\makebox(0,0)[c]{8}}
\put(6.5,3.5){\makebox(0,0)[c]{7}}
\put(0.5,4.5){\makebox(0,0)[c]{0}}
\put(1.5,4.5){\makebox(0,0)[c]{1}}
\put(2.5,4.5){\makebox(0,0)[c]{0}}
\put(3.5,4.5){\makebox(0,0)[c]{5}}
\put(4.5,4.5){\makebox(0,0)[c]{8}}
\put(5.5,4.5){\makebox(0,0)[c]{7}}
\put(6.5,4.5){\makebox(0,0)[c]{8}}
\put(0.5,5.5){\makebox(0,0)[c]{1}}
\put(1.5,5.5){\makebox(0,0)[c]{0}}
\put(2.5,5.5){\makebox(0,0)[c]{1}}
\put(3.5,5.5){\makebox(0,0)[c]{6}}
\put(4.5,5.5){\makebox(0,0)[c]{7}}
\put(5.5,5.5){\makebox(0,0)[c]{8}}
\put(6.5,5.5){\makebox(0,0)[c]{7}}
\put(0.5,6.5){\makebox(0,0)[c]{0}}
\put(1.5,6.5){\makebox(0,0)[c]{1}}
\put(2.5,6.5){\makebox(0,0)[c]{0}}
\put(3.5,6.5){\makebox(0,0)[c]{7}}
\put(4.5,6.5){\makebox(0,0)[c]{8}}
\put(5.5,6.5){\makebox(0,0)[c]{7}}
\put(6.5,6.5){\makebox(0,0)[c]{8}}
\end{picture}
\hspace*{\fill}
\setlength{\unitlength}{0.5cm}
\begin{picture}(7,7)(0,0)
\setcoordinatesystem units <0.5cm,0.5cm> point at 0 0
\setplotarea x from 0 to 7, y from 0 to 7
\put(0.5,0.5){\makebox(0,0)[c]{0}}
\put(0.5,0.5){\circle{0.8}}
\put(1.5,0.5){\makebox(0,0)[c]{1}}
\put(1.5,0.5){\circle{0.8}}
\put(2.5,0.5){\makebox(0,0)[c]{0}}
\put(2.5,0.5){\circle{0.8}}
\put(3.5,0.5){\makebox(0,0)[c]{1}}
\put(3.5,0.5){\circle{0.8}}
\put(4.5,0.5){\makebox(0,0)[c]{8}}
\put(4.5,0.5){\circle{0.8}}
\put(5.5,0.5){\makebox(0,0)[c]{7}}
\put(5.5,0.5){\circle{0.8}}
\put(6.5,0.5){\makebox(0,0)[c]{8}}
\put(6.5,0.5){\circle{0.8}}
\put(0.5,1.5){\makebox(0,0)[c]{1}}
\put(0.5,1.5){\circle{0.8}}
\put(1.5,1.5){\makebox(0,0)[c]{0}}
\put(1.5,1.5){\circle{0.8}}
\put(2.5,1.5){\makebox(0,0)[c]{1}}
\put(2.5,1.5){\circle{0.8}}
\put(3.5,1.5){\makebox(0,0)[c]{2}}
\put(3.5,1.5){\circle{0.8}}
\put(4.5,1.5){\makebox(0,0)[c]{7}}
\put(4.5,1.5){\circle{0.8}}
\put(5.5,1.5){\makebox(0,0)[c]{8}}
\put(5.5,1.5){\circle{0.8}}
\put(6.5,1.5){\makebox(0,0)[c]{7}}
\put(6.5,1.5){\circle{0.8}}
\put(0.5,2.5){\makebox(0,0)[c]{0}}
\put(0.5,2.5){\circle{0.8}}
\put(1.5,2.5){\makebox(0,0)[c]{1}}
\put(1.5,2.5){\circle{0.8}}
\put(2.5,2.5){\makebox(0,0)[c]{0}}
\put(2.5,2.5){\circle{0.8}}
\put(3.5,2.5){\makebox(0,0)[c]{3}}
\put(3.5,2.5){\circle{0.8}}
\put(4.5,2.5){\makebox(0,0)[c]{8}}
\put(4.5,2.5){\circle{0.8}}
\put(5.5,2.5){\makebox(0,0)[c]{7}}
\put(5.5,2.5){\circle{0.8}}
\put(6.5,2.5){\makebox(0,0)[c]{8}}
\put(6.5,2.5){\circle{0.8}}
\put(0.5,3.5){\makebox(0,0)[c]{1}}
\put(0.5,3.5){\circle{0.8}}
\put(1.5,3.5){\makebox(0,0)[c]{0}}
\put(1.5,3.5){\circle{0.8}}
\put(2.5,3.5){\makebox(0,0)[c]{1}}
\put(2.5,3.5){\circle{0.8}}
\put(3.5,3.5){\makebox(0,0)[c]{4}}
\put(3.5,3.5){\circle{0.8}}
\put(4.5,3.5){\makebox(0,0)[c]{7}}
\put(4.5,3.5){\circle{0.8}}
\put(5.5,3.5){\makebox(0,0)[c]{8}}
\put(5.5,3.5){\circle{0.8}}
\put(6.5,3.5){\makebox(0,0)[c]{7}}
\put(6.5,3.5){\circle{0.8}}
\put(0.5,4.5){\makebox(0,0)[c]{0}}
\put(0.5,4.5){\circle{0.8}}
\put(1.5,4.5){\makebox(0,0)[c]{1}}
\put(1.5,4.5){\circle{0.8}}
\put(2.5,4.5){\makebox(0,0)[c]{0}}
\put(2.5,4.5){\circle{0.8}}
\put(3.5,4.5){\makebox(0,0)[c]{5}}
\put(3.5,4.5){\circle{0.8}}
\put(4.5,4.5){\makebox(0,0)[c]{8}}
\put(4.5,4.5){\circle{0.8}}
\put(5.5,4.5){\makebox(0,0)[c]{7}}
\put(5.5,4.5){\circle{0.8}}
\put(6.5,4.5){\makebox(0,0)[c]{8}}
\put(6.5,4.5){\circle{0.8}}
\put(0.5,5.5){\makebox(0,0)[c]{1}}
\put(0.5,5.5){\circle{0.8}}
\put(1.5,5.5){\makebox(0,0)[c]{0}}
\put(1.5,5.5){\circle{0.8}}
\put(2.5,5.5){\makebox(0,0)[c]{1}}
\put(2.5,5.5){\circle{0.8}}
\put(3.5,5.5){\makebox(0,0)[c]{6}}
\put(3.5,5.5){\circle{0.8}}
\put(4.5,5.5){\makebox(0,0)[c]{7}}
\put(4.5,5.5){\circle{0.8}}
\put(5.5,5.5){\makebox(0,0)[c]{8}}
\put(5.5,5.5){\circle{0.8}}
\put(6.5,5.5){\makebox(0,0)[c]{7}}
\put(6.5,5.5){\circle{0.8}}
\put(0.5,6.5){\makebox(0,0)[c]{0}}
\put(0.5,6.5){\circle{0.8}}
\put(1.5,6.5){\makebox(0,0)[c]{1}}
\put(1.5,6.5){\circle{0.8}}
\put(2.5,6.5){\makebox(0,0)[c]{0}}
\put(2.5,6.5){\circle{0.8}}
\put(3.5,6.5){\makebox(0,0)[c]{7}}
\put(3.5,6.5){\circle{0.8}}
\put(4.5,6.5){\makebox(0,0)[c]{8}}
\put(4.5,6.5){\circle{0.8}}
\put(5.5,6.5){\makebox(0,0)[c]{7}}
\put(5.5,6.5){\circle{0.8}}
\put(6.5,6.5){\makebox(0,0)[c]{8}}
\put(6.5,6.5){\circle{0.8}}
\allinethickness{1.5pt}
\put(0,0){\line(1,0){7}}
\put(0,0){\line(0,1){7}}
\put(7,7){\line(-1,0){7}}
\put(7,7){\line(0,-1){7}}
\put(1,7){\line(0,-1){1}}
\put(2,7){\line(0,-1){1}}
\put(3,7){\line(0,-1){1}}
\put(4,7){\line(0,-1){1}}
\put(5,7){\line(0,-1){1}}
\put(6,7){\line(0,-1){1}}
\put(1,6){\line(0,-1){1}}
\put(2,6){\line(0,-1){1}}
\put(3,6){\line(0,-1){1}}
\put(4,6){\line(0,-1){1}}
\put(5,6){\line(0,-1){1}}
\put(6,6){\line(0,-1){1}}
\put(1,5){\line(0,-1){1}}
\put(2,5){\line(0,-1){1}}
\put(3,5){\line(0,-1){1}}
\put(4,5){\line(0,-1){1}}
\put(5,5){\line(0,-1){1}}
\put(6,5){\line(0,-1){1}}
\put(1,4){\line(0,-1){1}}
\put(2,4){\line(0,-1){1}}
\put(3,4){\line(0,-1){1}}
\put(4,4){\line(0,-1){1}}
\put(5,4){\line(0,-1){1}}
\put(6,4){\line(0,-1){1}}
\put(1,3){\line(0,-1){1}}
\put(2,3){\line(0,-1){1}}
\put(3,3){\line(0,-1){1}}
\put(4,3){\line(0,-1){1}}
\put(5,3){\line(0,-1){1}}
\put(6,3){\line(0,-1){1}}
\put(1,2){\line(0,-1){1}}
\put(2,2){\line(0,-1){1}}
\put(3,2){\line(0,-1){1}}
\put(4,2){\line(0,-1){1}}
\put(5,2){\line(0,-1){1}}
\put(6,2){\line(0,-1){1}}
\put(1,1){\line(0,-1){1}}
\put(2,1){\line(0,-1){1}}
\put(3,1){\line(0,-1){1}}
\put(4,1){\line(0,-1){1}}
\put(5,1){\line(0,-1){1}}
\put(6,1){\line(0,-1){1}}
\put(0,6){\line(1,0){1}}
\put(0,5){\line(1,0){1}}
\put(0,4){\line(1,0){1}}
\put(0,3){\line(1,0){1}}
\put(0,2){\line(1,0){1}}
\put(0,1){\line(1,0){1}}
\put(1,6){\line(1,0){1}}
\put(1,5){\line(1,0){1}}
\put(1,4){\line(1,0){1}}
\put(1,3){\line(1,0){1}}
\put(1,2){\line(1,0){1}}
\put(1,1){\line(1,0){1}}
\put(2,6){\line(1,0){1}}
\put(2,5){\line(1,0){1}}
\put(2,4){\line(1,0){1}}
\put(2,3){\line(1,0){1}}
\put(2,2){\line(1,0){1}}
\put(2,1){\line(1,0){1}}
\put(3,6){\line(1,0){1}}
\put(3,5){\line(1,0){1}}
\put(3,4){\line(1,0){1}}
\put(3,3){\line(1,0){1}}
\put(3,2){\line(1,0){1}}
\put(3,1){\line(1,0){1}}
\put(4,6){\line(1,0){1}}
\put(4,5){\line(1,0){1}}
\put(4,4){\line(1,0){1}}
\put(4,3){\line(1,0){1}}
\put(4,2){\line(1,0){1}}
\put(4,1){\line(1,0){1}}
\put(5,6){\line(1,0){1}}
\put(5,5){\line(1,0){1}}
\put(5,4){\line(1,0){1}}
\put(5,3){\line(1,0){1}}
\put(5,2){\line(1,0){1}}
\put(5,1){\line(1,0){1}}
\put(6,6){\line(1,0){1}}
\put(6,5){\line(1,0){1}}
\put(6,4){\line(1,0){1}}
\put(6,3){\line(1,0){1}}
\put(6,2){\line(1,0){1}}
\put(6,1){\line(1,0){1}}
\end{picture}
\hspace*{\fill}
\setlength{\unitlength}{0.5cm}
\begin{picture}(7,7)(0,0)
\setcoordinatesystem units <0.5cm,0.5cm> point at 0 0
\setplotarea x from 0 to 7, y from 0 to 7
\put(0.5,0.5){\makebox(0,0)[c]{0}}
\put(0.5,0.5){\circle{0.8}}
\put(1.5,0.5){\makebox(0,0)[c]{1}}
\put(1.5,0.5){\circle{0.8}}
\put(2.5,0.5){\makebox(0,0)[c]{0}}
\put(2.5,0.5){\circle{0.8}}
\put(3.5,0.5){\makebox(0,0)[c]{1}}
\put(3.5,0.5){\circle{0.8}}
\put(4.5,0.5){\makebox(0,0)[c]{8}}
\put(4.5,0.5){\circle{0.8}}
\put(5.5,0.5){\makebox(0,0)[c]{7}}
\put(5.5,0.5){\circle{0.8}}
\put(6.5,0.5){\makebox(0,0)[c]{8}}
\put(6.5,0.5){\circle{0.8}}
\put(0.5,1.5){\makebox(0,0)[c]{1}}
\put(0.5,1.5){\circle{0.8}}
\put(1.5,1.5){\makebox(0,0)[c]{0}}
\put(1.5,1.5){\circle{0.8}}
\put(2.5,1.5){\makebox(0,0)[c]{1}}
\put(2.5,1.5){\circle{0.8}}
\put(3.5,1.5){\makebox(0,0)[c]{2}}
\put(3.5,1.5){\circle{0.8}}
\put(4.5,1.5){\makebox(0,0)[c]{7}}
\put(4.5,1.5){\circle{0.8}}
\put(5.5,1.5){\makebox(0,0)[c]{8}}
\put(5.5,1.5){\circle{0.8}}
\put(6.5,1.5){\makebox(0,0)[c]{7}}
\put(6.5,1.5){\circle{0.8}}
\put(0.5,2.5){\makebox(0,0)[c]{0}}
\put(0.5,2.5){\circle{0.8}}
\put(1.5,2.5){\makebox(0,0)[c]{1}}
\put(1.5,2.5){\circle{0.8}}
\put(2.5,2.5){\makebox(0,0)[c]{0}}
\put(2.5,2.5){\circle{0.8}}
\put(3.5,2.5){\makebox(0,0)[c]{3}}
\put(3.5,2.5){\circle{0.8}}
\put(4.5,2.5){\makebox(0,0)[c]{8}}
\put(4.5,2.5){\circle{0.8}}
\put(5.5,2.5){\makebox(0,0)[c]{7}}
\put(5.5,2.5){\circle{0.8}}
\put(6.5,2.5){\makebox(0,0)[c]{8}}
\put(6.5,2.5){\circle{0.8}}
\put(0.5,3.5){\makebox(0,0)[c]{1}}
\put(0.5,3.5){\circle{0.8}}
\put(1.5,3.5){\makebox(0,0)[c]{0}}
\put(1.5,3.5){\circle{0.8}}
\put(2.5,3.5){\makebox(0,0)[c]{1}}
\put(2.5,3.5){\circle{0.8}}
\put(3.5,3.5){\makebox(0,0)[c]{4}}
\put(3.5,3.5){\circle{0.8}}
\put(4.5,3.5){\makebox(0,0)[c]{7}}
\put(4.5,3.5){\circle{0.8}}
\put(5.5,3.5){\makebox(0,0)[c]{8}}
\put(5.5,3.5){\circle{0.8}}
\put(6.5,3.5){\makebox(0,0)[c]{7}}
\put(6.5,3.5){\circle{0.8}}
\put(0.5,4.5){\makebox(0,0)[c]{0}}
\put(0.5,4.5){\circle{0.8}}
\put(1.5,4.5){\makebox(0,0)[c]{1}}
\put(1.5,4.5){\circle{0.8}}
\put(2.5,4.5){\makebox(0,0)[c]{0}}
\put(2.5,4.5){\circle{0.8}}
\put(3.5,4.5){\makebox(0,0)[c]{5}}
\put(3.5,4.5){\circle{0.8}}
\put(4.5,4.5){\makebox(0,0)[c]{8}}
\put(4.5,4.5){\circle{0.8}}
\put(5.5,4.5){\makebox(0,0)[c]{7}}
\put(5.5,4.5){\circle{0.8}}
\put(6.5,4.5){\makebox(0,0)[c]{8}}
\put(6.5,4.5){\circle{0.8}}
\put(0.5,5.5){\makebox(0,0)[c]{1}}
\put(0.5,5.5){\circle{0.8}}
\put(1.5,5.5){\makebox(0,0)[c]{0}}
\put(1.5,5.5){\circle{0.8}}
\put(2.5,5.5){\makebox(0,0)[c]{1}}
\put(2.5,5.5){\circle{0.8}}
\put(3.5,5.5){\makebox(0,0)[c]{6}}
\put(3.5,5.5){\circle{0.8}}
\put(4.5,5.5){\makebox(0,0)[c]{7}}
\put(4.5,5.5){\circle{0.8}}
\put(5.5,5.5){\makebox(0,0)[c]{8}}
\put(5.5,5.5){\circle{0.8}}
\put(6.5,5.5){\makebox(0,0)[c]{7}}
\put(6.5,5.5){\circle{0.8}}
\put(0.5,6.5){\makebox(0,0)[c]{0}}
\put(0.5,6.5){\circle{0.8}}
\put(1.5,6.5){\makebox(0,0)[c]{1}}
\put(1.5,6.5){\circle{0.8}}
\put(2.5,6.5){\makebox(0,0)[c]{0}}
\put(2.5,6.5){\circle{0.8}}
\put(3.5,6.5){\makebox(0,0)[c]{7}}
\put(3.5,6.5){\circle{0.8}}
\put(4.5,6.5){\makebox(0,0)[c]{8}}
\put(4.5,6.5){\circle{0.8}}
\put(5.5,6.5){\makebox(0,0)[c]{7}}
\put(5.5,6.5){\circle{0.8}}
\put(6.5,6.5){\makebox(0,0)[c]{8}}
\put(6.5,6.5){\circle{0.8}}
\allinethickness{1.5pt}
\put(0,0){\line(1,0){7}}
\put(0,0){\line(0,1){7}}
\put(7,7){\line(-1,0){7}}
\put(7,7){\line(0,-1){7}}
\put(1,6.5){\line(1,0){0.11}}
\put(1,6.5){\line(-1,0){0.11}}
\put(2,6.5){\line(1,0){0.11}}
\put(2,6.5){\line(-1,0){0.11}}
\put(4,6.5){\line(1,0){0.11}}
\put(4,6.5){\line(-1,0){0.11}}
\put(5,6.5){\line(1,0){0.11}}
\put(5,6.5){\line(-1,0){0.11}}
\put(6,6.5){\line(1,0){0.11}}
\put(6,6.5){\line(-1,0){0.11}}
\put(1,5.5){\line(1,0){0.11}}
\put(1,5.5){\line(-1,0){0.11}}
\put(2,5.5){\line(1,0){0.11}}
\put(2,5.5){\line(-1,0){0.11}}
\put(4,5.5){\line(1,0){0.11}}
\put(4,5.5){\line(-1,0){0.11}}
\put(5,5.5){\line(1,0){0.11}}
\put(5,5.5){\line(-1,0){0.11}}
\put(6,5.5){\line(1,0){0.11}}
\put(6,5.5){\line(-1,0){0.11}}
\put(1,4.5){\line(1,0){0.11}}
\put(1,4.5){\line(-1,0){0.11}}
\put(2,4.5){\line(1,0){0.11}}
\put(2,4.5){\line(-1,0){0.11}}
\put(5,4.5){\line(1,0){0.11}}
\put(5,4.5){\line(-1,0){0.11}}
\put(6,4.5){\line(1,0){0.11}}
\put(6,4.5){\line(-1,0){0.11}}
\put(1,3.5){\line(1,0){0.11}}
\put(1,3.5){\line(-1,0){0.11}}
\put(2,3.5){\line(1,0){0.11}}
\put(2,3.5){\line(-1,0){0.11}}
\put(5,3.5){\line(1,0){0.11}}
\put(5,3.5){\line(-1,0){0.11}}
\put(6,3.5){\line(1,0){0.11}}
\put(6,3.5){\line(-1,0){0.11}}
\put(1,2.5){\line(1,0){0.11}}
\put(1,2.5){\line(-1,0){0.11}}
\put(2,2.5){\line(1,0){0.11}}
\put(2,2.5){\line(-1,0){0.11}}
\put(5,2.5){\line(1,0){0.11}}
\put(5,2.5){\line(-1,0){0.11}}
\put(6,2.5){\line(1,0){0.11}}
\put(6,2.5){\line(-1,0){0.11}}
\put(1,1.5){\line(1,0){0.11}}
\put(1,1.5){\line(-1,0){0.11}}
\put(2,1.5){\line(1,0){0.11}}
\put(2,1.5){\line(-1,0){0.11}}
\put(3,1.5){\line(1,0){0.11}}
\put(3,1.5){\line(-1,0){0.11}}
\put(5,1.5){\line(1,0){0.11}}
\put(5,1.5){\line(-1,0){0.11}}
\put(6,1.5){\line(1,0){0.11}}
\put(6,1.5){\line(-1,0){0.11}}
\put(1,0.5){\line(1,0){0.11}}
\put(1,0.5){\line(-1,0){0.11}}
\put(2,0.5){\line(1,0){0.11}}
\put(2,0.5){\line(-1,0){0.11}}
\put(3,0.5){\line(1,0){0.11}}
\put(3,0.5){\line(-1,0){0.11}}
\put(5,0.5){\line(1,0){0.11}}
\put(5,0.5){\line(-1,0){0.11}}
\put(6,0.5){\line(1,0){0.11}}
\put(6,0.5){\line(-1,0){0.11}}
\put(0.5,6){\line(0,1){0.11}}
\put(0.5,6){\line(0,-1){0.11}}
\put(0.5,5){\line(0,1){0.11}}
\put(0.5,5){\line(0,-1){0.11}}
\put(0.5,4){\line(0,1){0.11}}
\put(0.5,4){\line(0,-1){0.11}}
\put(0.5,3){\line(0,1){0.11}}
\put(0.5,3){\line(0,-1){0.11}}
\put(0.5,2){\line(0,1){0.11}}
\put(0.5,2){\line(0,-1){0.11}}
\put(0.5,1){\line(0,1){0.11}}
\put(0.5,1){\line(0,-1){0.11}}
\put(1.5,6){\line(0,1){0.11}}
\put(1.5,6){\line(0,-1){0.11}}
\put(1.5,5){\line(0,1){0.11}}
\put(1.5,5){\line(0,-1){0.11}}
\put(1.5,4){\line(0,1){0.11}}
\put(1.5,4){\line(0,-1){0.11}}
\put(1.5,3){\line(0,1){0.11}}
\put(1.5,3){\line(0,-1){0.11}}
\put(1.5,2){\line(0,1){0.11}}
\put(1.5,2){\line(0,-1){0.11}}
\put(1.5,1){\line(0,1){0.11}}
\put(1.5,1){\line(0,-1){0.11}}
\put(2.5,6){\line(0,1){0.11}}
\put(2.5,6){\line(0,-1){0.11}}
\put(2.5,5){\line(0,1){0.11}}
\put(2.5,5){\line(0,-1){0.11}}
\put(2.5,4){\line(0,1){0.11}}
\put(2.5,4){\line(0,-1){0.11}}
\put(2.5,3){\line(0,1){0.11}}
\put(2.5,3){\line(0,-1){0.11}}
\put(2.5,2){\line(0,1){0.11}}
\put(2.5,2){\line(0,-1){0.11}}
\put(2.5,1){\line(0,1){0.11}}
\put(2.5,1){\line(0,-1){0.11}}
\put(3.5,6){\line(0,1){0.11}}
\put(3.5,6){\line(0,-1){0.11}}
\put(3.5,5){\line(0,1){0.11}}
\put(3.5,5){\line(0,-1){0.11}}
\put(3.5,4){\line(0,1){0.11}}
\put(3.5,4){\line(0,-1){0.11}}
\put(3.5,3){\line(0,1){0.11}}
\put(3.5,3){\line(0,-1){0.11}}
\put(3.5,2){\line(0,1){0.11}}
\put(3.5,2){\line(0,-1){0.11}}
\put(3.5,1){\line(0,1){0.11}}
\put(3.5,1){\line(0,-1){0.11}}
\put(4.5,6){\line(0,1){0.11}}
\put(4.5,6){\line(0,-1){0.11}}
\put(4.5,5){\line(0,1){0.11}}
\put(4.5,5){\line(0,-1){0.11}}
\put(4.5,4){\line(0,1){0.11}}
\put(4.5,4){\line(0,-1){0.11}}
\put(4.5,3){\line(0,1){0.11}}
\put(4.5,3){\line(0,-1){0.11}}
\put(4.5,2){\line(0,1){0.11}}
\put(4.5,2){\line(0,-1){0.11}}
\put(4.5,1){\line(0,1){0.11}}
\put(4.5,1){\line(0,-1){0.11}}
\put(5.5,6){\line(0,1){0.11}}
\put(5.5,6){\line(0,-1){0.11}}
\put(5.5,5){\line(0,1){0.11}}
\put(5.5,5){\line(0,-1){0.11}}
\put(5.5,4){\line(0,1){0.11}}
\put(5.5,4){\line(0,-1){0.11}}
\put(5.5,3){\line(0,1){0.11}}
\put(5.5,3){\line(0,-1){0.11}}
\put(5.5,2){\line(0,1){0.11}}
\put(5.5,2){\line(0,-1){0.11}}
\put(5.5,1){\line(0,1){0.11}}
\put(5.5,1){\line(0,-1){0.11}}
\put(6.5,6){\line(0,1){0.11}}
\put(6.5,6){\line(0,-1){0.11}}
\put(6.5,5){\line(0,1){0.11}}
\put(6.5,5){\line(0,-1){0.11}}
\put(6.5,4){\line(0,1){0.11}}
\put(6.5,4){\line(0,-1){0.11}}
\put(6.5,3){\line(0,1){0.11}}
\put(6.5,3){\line(0,-1){0.11}}
\put(6.5,2){\line(0,1){0.11}}
\put(6.5,2){\line(0,-1){0.11}}
\put(6.5,1){\line(0,1){0.11}}
\put(6.5,1){\line(0,-1){0.11}}
\end{picture}
\hspace*{\fill}
\caption{\label{f.ccissue} A synthetic sample image with its intensity
  values and its two possible partitions into constrained connected
  components {\em whatever\/} the considered constraints in case
  standard $\arangepsl$-connectivity is used in the definitions.  The
  two homogeneous regions show intensity variations of~1 level while
  the ramp between the two regions also proceeds by steps of~1
  intensity level.  In the image at the right, adjacent pixels are
  linked by an edge if and only if their range does not exceed~1.  }
\end{figure}
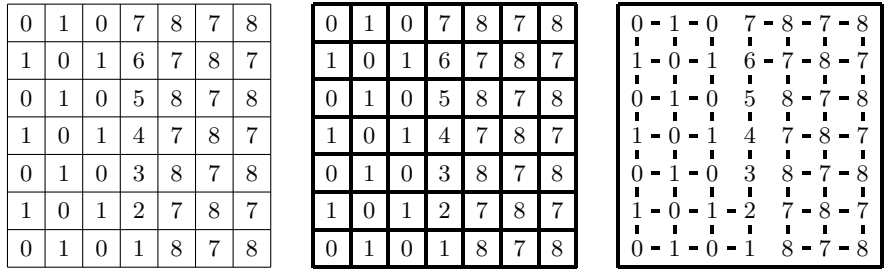

Constrained connectivity prevents the formation of connected
components that would otherwise be created in case samples of
intermediate value (transition pixels) between two populations
(homogeneous image structures) are present.  Indeed, these components
would violate the global range or other appropriate constraint.
However, sometimes the formation of two distinct connected components
cannot occur at all.  In the extreme case represented in
Fig.~\ref{f.ccissue}. either each pixel is a connected component (flat
zone) or there is a unique connected component.
One way to address this problem is to propose a definition of
transition pixels and perform some pre-processing to suppress them.
This approach is advocated
in~\citep{soille-grazzini2009ismm,soille2010ijrs}.  For example,
assuming that {\em local\/} extrema correspond to non-transition
pixels, they are extracted on then considered as seeds whose values
are propagated in the input image using a seeded region growing
algorithm~\citep{adams-bischof94}.  Note that this approach is linked
with contrast enhancement techniques since it aims at increasing the
external isolation of the obtained connected components.  A number of
classical morphological schemes ({\em e.g.}, area filtering of the
ultrametric watershed) can be used to remove those transition zones
(see Fig.~\ref{figX:alphaomega}.d for an example).
 
\begin{figure}
\hspace*{\fill}\setlength{\unitlength}{0.5cm}
\begin{picture}(7,7)(0,0)
\setcoordinatesystem units <0.5cm,0.5cm> point at 0 0
\setplotarea x from 0 to 7, y from 0 to 7
\grid {7} {7}
\put(0.5,0.5){\makebox(0,0)[c]{0}}
\put(1.5,0.5){\makebox(0,0)[c]{1}}
\put(2.5,0.5){\makebox(0,0)[c]{0}}
\put(3.5,0.5){\makebox(0,0)[c]{1}}
\put(4.5,0.5){\makebox(0,0)[c]{8}}
\put(5.5,0.5){\makebox(0,0)[c]{7}}
\put(6.5,0.5){\makebox(0,0)[c]{8}}
\put(0.5,1.5){\makebox(0,0)[c]{1}}
\put(1.5,1.5){\makebox(0,0)[c]{0}}
\put(2.5,1.5){\makebox(0,0)[c]{1}}
\put(3.5,1.5){\makebox(0,0)[c]{2}}
\put(4.5,1.5){\makebox(0,0)[c]{7}}
\put(5.5,1.5){\makebox(0,0)[c]{8}}
\put(6.5,1.5){\makebox(0,0)[c]{7}}
\put(0.5,2.5){\makebox(0,0)[c]{0}}
\put(1.5,2.5){\makebox(0,0)[c]{1}}
\put(2.5,2.5){\makebox(0,0)[c]{0}}
\put(3.5,2.5){\makebox(0,0)[c]{3}}
\put(4.5,2.5){\makebox(0,0)[c]{8}}
\put(5.5,2.5){\makebox(0,0)[c]{7}}
\put(6.5,2.5){\makebox(0,0)[c]{8}}
\put(0.5,3.5){\makebox(0,0)[c]{1}}
\put(1.5,3.5){\makebox(0,0)[c]{0}}
\put(2.5,3.5){\makebox(0,0)[c]{1}}
\put(3.5,3.5){\makebox(0,0)[c]{4}}
\put(4.5,3.5){\makebox(0,0)[c]{7}}
\put(5.5,3.5){\makebox(0,0)[c]{8}}
\put(6.5,3.5){\makebox(0,0)[c]{7}}
\put(0.5,4.5){\makebox(0,0)[c]{0}}
\put(1.5,4.5){\makebox(0,0)[c]{1}}
\put(2.5,4.5){\makebox(0,0)[c]{0}}
\put(3.5,4.5){\makebox(0,0)[c]{5}}
\put(4.5,4.5){\makebox(0,0)[c]{8}}
\put(5.5,4.5){\makebox(0,0)[c]{7}}
\put(6.5,4.5){\makebox(0,0)[c]{8}}
\put(0.5,5.5){\makebox(0,0)[c]{1}}
\put(1.5,5.5){\makebox(0,0)[c]{0}}
\put(2.5,5.5){\makebox(0,0)[c]{1}}
\put(3.5,5.5){\makebox(0,0)[c]{6}}
\put(4.5,5.5){\makebox(0,0)[c]{7}}
\put(5.5,5.5){\makebox(0,0)[c]{8}}
\put(6.5,5.5){\makebox(0,0)[c]{7}}
\put(0.5,6.5){\makebox(0,0)[c]{0}}
\put(1.5,6.5){\makebox(0,0)[c]{1}}
\put(2.5,6.5){\makebox(0,0)[c]{0}}
\put(3.5,6.5){\makebox(0,0)[c]{7}}
\put(4.5,6.5){\makebox(0,0)[c]{8}}
\put(5.5,6.5){\makebox(0,0)[c]{7}}
\put(6.5,6.5){\makebox(0,0)[c]{8}}
\end{picture}
\hspace*{\fill}
\setlength{\unitlength}{0.5cm}
\begin{picture}(7,7)(0,0)
\setcoordinatesystem units <0.5cm,0.5cm> point at 0 0
\setplotarea x from 0 to 7, y from 0 to 7
\grid {7} {7}
\put(0.5,0.5){\makebox(0,0)[c]{2}}
\put(1.5,0.5){\makebox(0,0)[c]{3}}
\put(2.5,0.5){\makebox(0,0)[c]{3}}
\put(3.5,0.5){\makebox(0,0)[c]{2}}
\put(4.5,0.5){\makebox(0,0)[c]{2}}
\put(5.5,0.5){\makebox(0,0)[c]{3}}
\put(6.5,0.5){\makebox(0,0)[c]{2}}
\put(0.5,1.5){\makebox(0,0)[c]{3}}
\put(1.5,1.5){\makebox(0,0)[c]{4}}
\put(2.5,1.5){\makebox(0,0)[c]{4}}
\put(3.5,1.5){\makebox(0,0)[c]{3}}
\put(4.5,1.5){\makebox(0,0)[c]{3}}
\put(5.5,1.5){\makebox(0,0)[c]{4}}
\put(6.5,1.5){\makebox(0,0)[c]{3}}
\put(0.5,2.5){\makebox(0,0)[c]{3}}
\put(1.5,2.5){\makebox(0,0)[c]{4}}
\put(2.5,2.5){\makebox(0,0)[c]{3}}
\put(3.5,2.5){\makebox(0,0)[c]{2}}
\put(4.5,2.5){\makebox(0,0)[c]{3}}
\put(5.5,2.5){\makebox(0,0)[c]{4}}
\put(6.5,2.5){\makebox(0,0)[c]{3}}
\put(0.5,3.5){\makebox(0,0)[c]{3}}
\put(1.5,3.5){\makebox(0,0)[c]{4}}
\put(2.5,3.5){\makebox(0,0)[c]{3}}
\put(3.5,3.5){\makebox(0,0)[c]{2}}
\put(4.5,3.5){\makebox(0,0)[c]{3}}
\put(5.5,3.5){\makebox(0,0)[c]{4}}
\put(6.5,3.5){\makebox(0,0)[c]{3}}
\put(0.5,4.5){\makebox(0,0)[c]{3}}
\put(1.5,4.5){\makebox(0,0)[c]{4}}
\put(2.5,4.5){\makebox(0,0)[c]{3}}
\put(3.5,4.5){\makebox(0,0)[c]{2}}
\put(4.5,4.5){\makebox(0,0)[c]{3}}
\put(5.5,4.5){\makebox(0,0)[c]{4}}
\put(6.5,4.5){\makebox(0,0)[c]{3}}
\put(0.5,5.5){\makebox(0,0)[c]{3}}
\put(1.5,5.5){\makebox(0,0)[c]{4}}
\put(2.5,5.5){\makebox(0,0)[c]{3}}
\put(3.5,5.5){\makebox(0,0)[c]{3}}
\put(4.5,5.5){\makebox(0,0)[c]{4}}
\put(5.5,5.5){\makebox(0,0)[c]{4}}
\put(6.5,5.5){\makebox(0,0)[c]{3}}
\put(0.5,6.5){\makebox(0,0)[c]{2}}
\put(1.5,6.5){\makebox(0,0)[c]{3}}
\put(2.5,6.5){\makebox(0,0)[c]{2}}
\put(3.5,6.5){\makebox(0,0)[c]{2}}
\put(4.5,6.5){\makebox(0,0)[c]{3}}
\put(5.5,6.5){\makebox(0,0)[c]{3}}
\put(6.5,6.5){\makebox(0,0)[c]{2}}
\end{picture}
\hspace*{\fill}
\setlength{\unitlength}{0.5cm}
\begin{picture}(7,7)(0,0)
\setcoordinatesystem units <0.5cm,0.5cm> point at 0 0
\setplotarea x from 0 to 7, y from 0 to 7
\put(0.5,0.5){\makebox(0,0)[c]{0}}
\put(0.5,0.5){\circle{0.8}}
\put(1.5,0.5){\makebox(0,0)[c]{1}}
\put(1.5,0.5){\circle{0.8}}
\put(2.5,0.5){\makebox(0,0)[c]{0}}
\put(2.5,0.5){\circle{0.8}}
\put(3.5,0.5){\makebox(0,0)[c]{1}}
\put(3.5,0.5){\circle{0.8}}
\put(4.5,0.5){\makebox(0,0)[c]{8}}
\put(4.5,0.5){\circle{0.8}}
\put(5.5,0.5){\makebox(0,0)[c]{7}}
\put(5.5,0.5){\circle{0.8}}
\put(6.5,0.5){\makebox(0,0)[c]{8}}
\put(6.5,0.5){\circle{0.8}}
\put(0.5,1.5){\makebox(0,0)[c]{1}}
\put(0.5,1.5){\circle{0.8}}
\put(1.5,1.5){\makebox(0,0)[c]{0}}
\put(1.5,1.5){\circle{0.8}}
\put(2.5,1.5){\makebox(0,0)[c]{1}}
\put(2.5,1.5){\circle{0.8}}
\put(3.5,1.5){\makebox(0,0)[c]{2}}
\put(3.5,1.5){\circle{0.8}}
\put(4.5,1.5){\makebox(0,0)[c]{7}}
\put(4.5,1.5){\circle{0.8}}
\put(5.5,1.5){\makebox(0,0)[c]{8}}
\put(5.5,1.5){\circle{0.8}}
\put(6.5,1.5){\makebox(0,0)[c]{7}}
\put(6.5,1.5){\circle{0.8}}
\put(0.5,2.5){\makebox(0,0)[c]{0}}
\put(0.5,2.5){\circle{0.8}}
\put(1.5,2.5){\makebox(0,0)[c]{1}}
\put(1.5,2.5){\circle{0.8}}
\put(2.5,2.5){\makebox(0,0)[c]{0}}
\put(2.5,2.5){\circle{0.8}}
\put(3.5,2.5){\makebox(0,0)[c]{3}}
\put(3.5,2.5){\circle{0.8}}
\put(4.5,2.5){\makebox(0,0)[c]{8}}
\put(4.5,2.5){\circle{0.8}}
\put(5.5,2.5){\makebox(0,0)[c]{7}}
\put(5.5,2.5){\circle{0.8}}
\put(6.5,2.5){\makebox(0,0)[c]{8}}
\put(6.5,2.5){\circle{0.8}}
\put(0.5,3.5){\makebox(0,0)[c]{1}}
\put(0.5,3.5){\circle{0.8}}
\put(1.5,3.5){\makebox(0,0)[c]{0}}
\put(1.5,3.5){\circle{0.8}}
\put(2.5,3.5){\makebox(0,0)[c]{1}}
\put(2.5,3.5){\circle{0.8}}
\put(3.5,3.5){\makebox(0,0)[c]{4}}
\put(3.5,3.5){\circle{0.8}}
\put(4.5,3.5){\makebox(0,0)[c]{7}}
\put(4.5,3.5){\circle{0.8}}
\put(5.5,3.5){\makebox(0,0)[c]{8}}
\put(5.5,3.5){\circle{0.8}}
\put(6.5,3.5){\makebox(0,0)[c]{7}}
\put(6.5,3.5){\circle{0.8}}
\put(0.5,4.5){\makebox(0,0)[c]{0}}
\put(0.5,4.5){\circle{0.8}}
\put(1.5,4.5){\makebox(0,0)[c]{1}}
\put(1.5,4.5){\circle{0.8}}
\put(2.5,4.5){\makebox(0,0)[c]{0}}
\put(2.5,4.5){\circle{0.8}}
\put(3.5,4.5){\makebox(0,0)[c]{5}}
\put(3.5,4.5){\circle{0.8}}
\put(4.5,4.5){\makebox(0,0)[c]{8}}
\put(4.5,4.5){\circle{0.8}}
\put(5.5,4.5){\makebox(0,0)[c]{7}}
\put(5.5,4.5){\circle{0.8}}
\put(6.5,4.5){\makebox(0,0)[c]{8}}
\put(6.5,4.5){\circle{0.8}}
\put(0.5,5.5){\makebox(0,0)[c]{1}}
\put(0.5,5.5){\circle{0.8}}
\put(1.5,5.5){\makebox(0,0)[c]{0}}
\put(1.5,5.5){\circle{0.8}}
\put(2.5,5.5){\makebox(0,0)[c]{1}}
\put(2.5,5.5){\circle{0.8}}
\put(3.5,5.5){\makebox(0,0)[c]{6}}
\put(3.5,5.5){\circle{0.8}}
\put(4.5,5.5){\makebox(0,0)[c]{7}}
\put(4.5,5.5){\circle{0.8}}
\put(5.5,5.5){\makebox(0,0)[c]{8}}
\put(5.5,5.5){\circle{0.8}}
\put(6.5,5.5){\makebox(0,0)[c]{7}}
\put(6.5,5.5){\circle{0.8}}
\put(0.5,6.5){\makebox(0,0)[c]{0}}
\put(0.5,6.5){\circle{0.8}}
\put(1.5,6.5){\makebox(0,0)[c]{1}}
\put(1.5,6.5){\circle{0.8}}
\put(2.5,6.5){\makebox(0,0)[c]{0}}
\put(2.5,6.5){\circle{0.8}}
\put(3.5,6.5){\makebox(0,0)[c]{7}}
\put(3.5,6.5){\circle{0.8}}
\put(4.5,6.5){\makebox(0,0)[c]{8}}
\put(4.5,6.5){\circle{0.8}}
\put(5.5,6.5){\makebox(0,0)[c]{7}}
\put(5.5,6.5){\circle{0.8}}
\put(6.5,6.5){\makebox(0,0)[c]{8}}
\put(6.5,6.5){\circle{0.8}}
\allinethickness{1.5pt}
\put(0,0){\line(1,0){7}}
\put(0,0){\line(0,1){7}}
\put(7,7){\line(-1,0){7}}
\put(7,7){\line(0,-1){7}}
\put(1,7){\line(0,-1){1}}
\put(2,7){\line(0,-1){1}}
\put(3,7){\line(0,-1){1}}
\put(4,7){\line(0,-1){1}}
\put(5,6.5){\line(1,0){0.11}}
\put(5,6.5){\line(-1,0){0.11}}
\put(6,7){\line(0,-1){1}}
\put(1,5.5){\line(1,0){0.11}}
\put(1,5.5){\line(-1,0){0.11}}
\put(2,5.5){\line(1,0){0.11}}
\put(2,5.5){\line(-1,0){0.11}}
\put(3,6){\line(0,-1){1}}
\put(4,5.5){\line(1,0){0.11}}
\put(4,5.5){\line(-1,0){0.11}}
\put(5,5.5){\line(1,0){0.11}}
\put(5,5.5){\line(-1,0){0.11}}
\put(6,5.5){\line(1,0){0.11}}
\put(6,5.5){\line(-1,0){0.11}}
\put(1,4.5){\line(1,0){0.11}}
\put(1,4.5){\line(-1,0){0.11}}
\put(2,4.5){\line(1,0){0.11}}
\put(2,4.5){\line(-1,0){0.11}}
\put(3,5){\line(0,-1){1}}
\put(4,5){\line(0,-1){1}}
\put(5,4.5){\line(1,0){0.11}}
\put(5,4.5){\line(-1,0){0.11}}
\put(6,4.5){\line(1,0){0.11}}
\put(6,4.5){\line(-1,0){0.11}}
\put(1,3.5){\line(1,0){0.11}}
\put(1,3.5){\line(-1,0){0.11}}
\put(2,3.5){\line(1,0){0.11}}
\put(2,3.5){\line(-1,0){0.11}}
\put(3,4){\line(0,-1){1}}
\put(4,4){\line(0,-1){1}}
\put(5,3.5){\line(1,0){0.11}}
\put(5,3.5){\line(-1,0){0.11}}
\put(6,3.5){\line(1,0){0.11}}
\put(6,3.5){\line(-1,0){0.11}}
\put(1,2.5){\line(1,0){0.11}}
\put(1,2.5){\line(-1,0){0.11}}
\put(2,2.5){\line(1,0){0.11}}
\put(2,2.5){\line(-1,0){0.11}}
\put(3,3){\line(0,-1){1}}
\put(4,3){\line(0,-1){1}}
\put(5,2.5){\line(1,0){0.11}}
\put(5,2.5){\line(-1,0){0.11}}
\put(6,2.5){\line(1,0){0.11}}
\put(6,2.5){\line(-1,0){0.11}}
\put(1,1.5){\line(1,0){0.11}}
\put(1,1.5){\line(-1,0){0.11}}
\put(2,1.5){\line(1,0){0.11}}
\put(2,1.5){\line(-1,0){0.11}}
\put(3,1.5){\line(1,0){0.11}}
\put(3,1.5){\line(-1,0){0.11}}
\put(4,2){\line(0,-1){1}}
\put(5,1.5){\line(1,0){0.11}}
\put(5,1.5){\line(-1,0){0.11}}
\put(6,1.5){\line(1,0){0.11}}
\put(6,1.5){\line(-1,0){0.11}}
\put(1,1){\line(0,-1){1}}
\put(2,0.5){\line(1,0){0.11}}
\put(2,0.5){\line(-1,0){0.11}}
\put(3,1){\line(0,-1){1}}
\put(4,1){\line(0,-1){1}}
\put(5,1){\line(0,-1){1}}
\put(6,1){\line(0,-1){1}}
\put(0,6){\line(1,0){1}}
\put(0.5,5){\line(0,1){0.11}}
\put(0.5,5){\line(0,-1){0.11}}
\put(0.5,4){\line(0,1){0.11}}
\put(0.5,4){\line(0,-1){0.11}}
\put(0.5,3){\line(0,1){0.11}}
\put(0.5,3){\line(0,-1){0.11}}
\put(0.5,2){\line(0,1){0.11}}
\put(0.5,2){\line(0,-1){0.11}}
\put(0,1){\line(1,0){1}}
\put(1.5,6){\line(0,1){0.11}}
\put(1.5,6){\line(0,-1){0.11}}
\put(1.5,5){\line(0,1){0.11}}
\put(1.5,5){\line(0,-1){0.11}}
\put(1.5,4){\line(0,1){0.11}}
\put(1.5,4){\line(0,-1){0.11}}
\put(1.5,3){\line(0,1){0.11}}
\put(1.5,3){\line(0,-1){0.11}}
\put(1.5,2){\line(0,1){0.11}}
\put(1.5,2){\line(0,-1){0.11}}
\put(1.5,1){\line(0,1){0.11}}
\put(1.5,1){\line(0,-1){0.11}}
\put(2,6){\line(1,0){1}}
\put(2.5,5){\line(0,1){0.11}}
\put(2.5,5){\line(0,-1){0.11}}
\put(2.5,4){\line(0,1){0.11}}
\put(2.5,4){\line(0,-1){0.11}}
\put(2.5,3){\line(0,1){0.11}}
\put(2.5,3){\line(0,-1){0.11}}
\put(2.5,2){\line(0,1){0.11}}
\put(2.5,2){\line(0,-1){0.11}}
\put(2.5,1){\line(0,1){0.11}}
\put(2.5,1){\line(0,-1){0.11}}
\put(3,6){\line(1,0){1}}
\put(3,5){\line(1,0){1}}
\put(3,4){\line(1,0){1}}
\put(3,3){\line(1,0){1}}
\put(3,2){\line(1,0){1}}
\put(3,1){\line(1,0){1}}
\put(4.5,6){\line(0,1){0.11}}
\put(4.5,6){\line(0,-1){0.11}}
\put(4.5,5){\line(0,1){0.11}}
\put(4.5,5){\line(0,-1){0.11}}
\put(4.5,4){\line(0,1){0.11}}
\put(4.5,4){\line(0,-1){0.11}}
\put(4.5,3){\line(0,1){0.11}}
\put(4.5,3){\line(0,-1){0.11}}
\put(4.5,2){\line(0,1){0.11}}
\put(4.5,2){\line(0,-1){0.11}}
\put(4,1){\line(1,0){1}}
\put(5.5,6){\line(0,1){0.11}}
\put(5.5,6){\line(0,-1){0.11}}
\put(5.5,5){\line(0,1){0.11}}
\put(5.5,5){\line(0,-1){0.11}}
\put(5.5,4){\line(0,1){0.11}}
\put(5.5,4){\line(0,-1){0.11}}
\put(5.5,3){\line(0,1){0.11}}
\put(5.5,3){\line(0,-1){0.11}}
\put(5.5,2){\line(0,1){0.11}}
\put(5.5,2){\line(0,-1){0.11}}
\put(5.5,1){\line(0,1){0.11}}
\put(5.5,1){\line(0,-1){0.11}}
\put(6,6){\line(1,0){1}}
\put(6.5,5){\line(0,1){0.11}}
\put(6.5,5){\line(0,-1){0.11}}
\put(6.5,4){\line(0,1){0.11}}
\put(6.5,4){\line(0,-1){0.11}}
\put(6.5,3){\line(0,1){0.11}}
\put(6.5,3){\line(0,-1){0.11}}
\put(6.5,2){\line(0,1){0.11}}
\put(6.5,2){\line(0,-1){0.11}}
\put(6,1){\line(1,0){1}}
\end{picture}
\hspace*{\fill}
\caption{\label{f.alphadeg} A synthetic sample image with its
  intensity values, the corresponding $\alphadegreeval{1}$ map, and
  $1_3$-connected components.}
\end{figure}
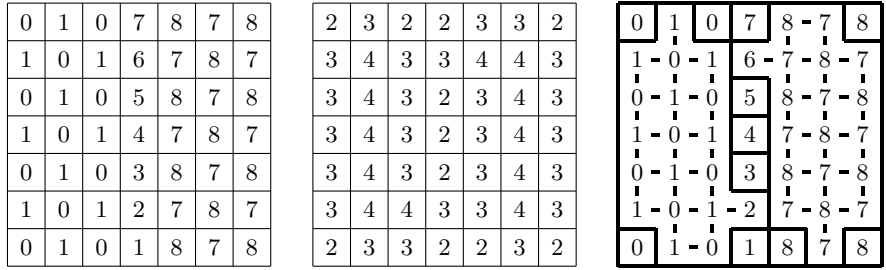

Another approach is to substitute the $\arangepsl$-connectivity with a
more restrictive connectivity.
Indeed, the local range parameter $\arangepsl$ defined
in~\citep{soille2008pami} as the intensity difference between adjacent
pixels can be viewed as a special case of dissimilarity measurement.
Although this measurement is the most natural, other dissimilarity
measurements may be considered.  For example, the following
alternative definition of alpha-connectivity may be considered to
tackle the problem of transition regions.  Let the $\arangepsl$-degree
of a pixel (node) be defined as the number of its adjacent pixels that
are within a range equal to~$\arangepsl$:
\[
\alphadegree(p)=\#\{\apixq\mid \{p,q\}\in E
\mbox{ and $|f(q)-f(p)|\leq \arangepsl$}\}.
\]
Then two pixels $\apixp$ and $\apixq$ are said to be
$\arangepsl_\alphaparam$-connected if and only if there exists an
$\arangepsl$-path connecting them such that every pixel of the path
has a $\arangepsl$-degree greater of equal to $\alphaparam$.  We
obtain therefore the following definition for the
$\arangepsl_\alphaparam$-connected component of a pixel $\apixp$:
\begin{eqnarray*}
\arangepsl_\alphaparam\RCC(\apixp)&=&
\{\apixp\}\cup \{\apixq\mid \mbox{ there exists a path $\langle\apixp=p_1,\ldots,p_n=\apixq\rangle$, $n>1$,}\\
&&\mbox{such that $|f(p_i)-f(p_{i+1}|\leq \arangepsl$ and $\alphadegree(p_i)\geq n$\}}.
\end{eqnarray*}
If necessary, other constraints can be considered.  Note that
$\arangepsl$-connectivity is a special case of
$\arangepsl_\alphaparam$-connecti\-vity obtained for $n=1$.  In
addition, the following nesting property holds:
\[
\arangepsl_{\alphaparam'}\RCC(p)\subseteq\arangepsl_{\alphaparam}\RCC(p),
\]
where $n\leq n'$.  $\arangepsl_\alphaparam$-connectivity satisfies all
properties of an equivalence relation and therefore also partitions
the image definition domain into unique maximal connected components.
An example is provided in Fig.~\ref{f.alphadeg}.
In this example, the non singleton $1_3$-connected components match
the core of the two homogeneous regions.  Singleton connected
components correspond to pixels whose degree is smaller than 3.
Non-singleton connected components can be used as seeds for coarsening
the obtained partition.  Special care is needed to produce connected
components matching one-pixel thick non-transition regions.  Alternative approaches to tackle the problem of transition regions are also presented in \citep{soille2011ismm} using a dissimilarity value taking into account the values of the gradient by erosion and dilation at the considered adjacent pixels and in \citep{gueguen-soille2011ismm} using image statistics.

\section{Conclusion and perspectives}\label{s.conclu}
In this paper, we have presented several equivalent tools dealing with
hierarchies of connected partitions. Such a review invites us to look
more closely at links between what have been done in different
research domains as, for example, between clustering and lattice
theory \citep{hubert72}. A first step in that direction is
\citep{IGMI_CouNajSer08}, and there is a need for in-depth study of
operators acting on lattices of graphs \citep{CouNajSer10} (or the one
of complexes \cite{IGMI_DiaCouNaj11}).  The question of transition
pixels is not only a theoretical one, regarding its significance for
applications. Finally, we want to stress the importance of having
frame work allowing a generic implementation of existing algorithms,
not limited to the pixel framework, but also able to deal
transparently with edges, or, more generally, with graphs and
complexes \citep{LevGerNaj12}.

Finally, when dealing with very large images such as those encountered in remote sensing or biomedical imaging, the computation of the min-tree of the edge graph of an image may be prohibitive in terms of memory needs (without mentioning the additional cost of doubling the graph to make sure that each flat zone of the original image is matched by a minimum of the edge graph).  In this situation, the direct computation of the alpha-tree of the image may be a valid alternative.  An efficient implementation based on the union-find as originally presented for the computation of component trees~\cite{NajCou2006} is presented in~\cite{ouzounis-soille2011alphatree}.                                                                                                                                                                                                                      


\bibliographystyle{splncs}
\end{document}